\documentclass[11pt]{article}

\usepackage{physics}

\usepackage{xcolor}
\usepackage{mathrsfs}

\usepackage{mathtools}

\usepackage{amsfonts}
\def\td{\tilde}
\newcommand{\hoch}[1]{$\, ^{#1}$}

\makeatletter
\@addtoreset{equation}{section}
\makeatother

\newcommand{\be}{\begin{equation}}
\newcommand{\ee}{\end{equation}}
\newcommand{\bea} {\begin{eqnarray}}
\newcommand{\eea}{\end{eqnarray}}
\newcommand{\nn}{\nonumber}

\newcommand{\overbar}[1]{\mkern 1.5mu\overline{\mkern-1.5mu#1\mkern-1.5mu}\mkern 1.5mu}

\def\ft#1#2{{\textstyle{\frac{\scriptstyle #1}{\scriptstyle #2} } }}
\def\fft#1#2{{\frac{#1}{#2}}}

\def\0{{\sst{(0)}}}
\def\1{{\sst{(1)}}}
\def\2{{\sst{(2)}}}
\def\3{{\sst{(3)}}}
\def\4{{\sst{(4)}}}
\def\5{{\sst{(5)}}}
\def\6{{\sst{(6)}}}
\def\7{{\sst{(7)}}}
\def\8{{\sst{(8)}}}
\def\sst#1{{\scriptscriptstyle #1}}
\def\oneone{\rlap 1\mkern4mu{\rm l}}
\def\ep{{\epsilon}}
\def\del{{\partial}}

\def\Hp{{{\wtd H}}}
\def\Hq{{{H}}}
\def\hp{{{\td h}}}
\def\hq{{{h}}}

\def\crampest{\medmuskip = 1mu plus 1mu minus 1mu}
\def\uncramp{\medmuskip = 4mu plus 2mu minus 4mu}

\def\cA{{{\cal A}}}
\def\cF{{{\cal F}}}
\def\tA{{{\widetilde A}}}
\def\tF{{{\widetilde F}}}

\def\cG{{{\cal G}}}

\def\bF{{{\overbar F}}}
\def\tbF{{{\widetilde\bF}}}

\def\tvp{{{\tilde\varphi}}}
\def\tchi{{{\widetilde\chi}}}

\def\mbI{{{\mathbb I}}}
\def\mbG{{{\mathbb G}}}
\def\mbB{{{\mathbb B}}}

\def\val{{{\vec\alpha\,}}}
\def\vbe{{{\vec\beta\,}}}
\def\vv{{{\vec v\,}}}
\def\cV{{{\cal V}}}
\def\cN{{{\cal N}}}
\def\cU{{{\cal U}}}

\def\im{{{\rm i\,}}}
\def\R{{\mathbb R}}

\def\wtd{\widetilde}

\def\bchi{{\bar\chi}}

\def\bQ{{{\overbar Q}}}
\def\bp{{{\bar p}}}

\def\ie{{ i.e.~}}

\def\Tr{{{\hbox{Tr}}}}

\begin{document}

\begin{flushright}
\hfill { UPR-1308-T\ \ \ MI-TH-212
}\\
\end{flushright}

\begin{center}
{\large {\bf  Conformal Symmetries  for 
Extremal Black Holes with General Asymptotic Scalars in STU Supergravity}}

\vspace{15pt}
{\large M. Cveti\v c$^{1,2}$, 
              C.N. Pope$^{3,4}$ and A. Saha$^{3}$}

\vspace{15pt}

{\hoch{1}\it Department of Physics and Astronomy, and Department of
 Mathematics,\\
University of Pennsylvania, Philadelphia, PA 19104, USA}

\vspace{10pt}

{\hoch{2}\it Center for Applied Mathematics and Theoretical Physics,\\
University of Maribor, SI2000 Maribor, Slovenia}

\vspace{10pt}

\hoch{3}{\it George P. \& Cynthia Woods Mitchell  Institute
for Fundamental Physics and Astronomy,\\
Texas A\&M University, College Station, TX 77843, USA}

\hoch{4}{\it DAMTP, Centre for Mathematical Sciences,
 Cambridge University,\\  Wilberforce Road, Cambridge CB3 OWA, UK}

\vspace{10pt}



\end{center}




\begin{abstract}

We present a construction of the most general BPS black holes of STU 
supergravity (${\cal N}=2$ supersymmetric $D=4$ supergravity coupled to 
three vector super-multiplets) with arbitrary asymptotic values of the scalar fields.  These solutions are obtained by acting with a subset of 
of the global symmetry generators on STU BPS black holes with zero 
values of the asymptotic scalars, both in the U-duality and 
the heterotic frame.  The solutions are parameterized by fourteen 
parameters:  four electric and four magnetic
 charges, and the asymptotic values of the six scalar fields. 
We also present BPS black hole solutions of a consistently truncated STU supergravity, which  are parameterized by two electric and two 
magnetic charges and two scalar fields. 
These latter solutions are significantly simplified, and are very 
suitable for further explicit studies. We also explore a conformal 
inversion symmetry of the Couch-Torrence type, which maps any member of the
fourteen-parameter family of BPS black holes to another member of the family.
Furthermore, these solutions are expected to be valuable in the studies 
of various swampland conjectures in the moduli space of 
string compactifications.

\end{abstract}

\pagebreak

\tableofcontents
\addtocontents{toc}{\protect\setcounter{tocdepth}{2}}



\def\half{\frac{1}{2}}
\def\ben{\begin{equation}}
\def\bea{\begin{eqnarray}}
\def\een{\end{equation}}
\def\eea{\end{eqnarray}}
\def \bv {{\bf v}}
\def \bs {{\bf s}}
\def\bt{{\bf t}}

\def \p {\partial}
\def \cL {{\cal  L}}
\def \cG {{\cal  G}}
\def \cLEG {{\cal LEG}}


\section{Introduction}

  Many intriguing non-perturbative aspects of string theory and M-theory 
have been brought to light by studying the black hole and higher $p$-brane 
solutions (see, for example, \cite{hulltown,witten}).  Especially
important in this context are the supersymmetric BPS solutions, which
are expected to be protected in the face of stringy
corrections to the leading-order effective action. Thus by
studying the BPS solutions in the low-energy supergravity limit one
can expect to gain insights that could remain relevant in the full
theory.

 The full four-dimensional supergravity theories resulting from the
dimensional reduction of the ten-dimensional heterotic or type II 
superstring are quite complicated, with many field strengths and scalar
fields in their bosonic sectors.  For many purposes, however, it
suffices to focus on the black hole solutions residing within a truncation
of the theories to the so-called STU supergravity, which comprises
${\cal N}=2$ supergravity coupled to three vector supermultiplets.  
Solutions within the STU theory can be rotated using the global 
symmetries of the full heterotic or type II theories to ``fill out''
solution sets of the larger theories. 

  There are different ways to present the STU supergravity, which are
related by dualisations of one or more of the four gauge field strengths. 
These different presentations of the theory arise naturally in 
different contexts.  For example, one of the duality complexions,
which we refer to as the heterotic formulation, is the one that arises
naturally when one performs a toroidal reduction from the ten-dimensional
heterotic theory and then makes a consistent truncation to the
STU supergravity. A different duality complexion, which we refer to
as the U-duality formulation, arises if one makes a consistent reduction
of eleven-dimensional supergravity on the 7-sphere, truncates further
to the subsector of fields that are invariant under the $U(1)$ maximal
torus of the $SO(8)$ isometry group of the 7-sphere, and then turns off
the gauge coupling by sending the radius of the sphere to infinity.  This 
formulation of STU supergravity is related to the heterotic formulation
by a dualisation of two out of the four field strengths.  There is also
an intermediate duality complexion which corresponds to dualising any
one of the four fields in the U-duality formulation.  We refer to this
as the 3+1 formulation of the theory. 

  As a preliminary to subsequent calculations, one of the purposes of
this paper is to obtain fully explicit relations between the fields in
the three above-mentioned formulations of the STU supergravity theory.
These will then be utilised later in the paper, when we construct
explicit expressions for the most general static BPS black hole
solutions in the STU supergravity theory.  The most general such solutions 
are characterised by a total of eight charges, corresponding to four
electric and four magnetic charges carried by the four gauge field 
strengths.  One can use the compact $U(1)^3$ subgroup of the 
$SL(2,\R)^3$ global symmetry of the STU theory to rotate these solutions to
ones involving only 5 independent charges, but for our purposes it is more
useful to present the solutions in the more symmetrical 8-charge
characterisation.  In fact, we obtain this symmetrical form by using
the $U(1)^3$ symmetry in the opposite direction, to go from a 5-charge to
an 8-charge expression.  This procedure was originally partially implemented
in \cite{cvettsey}; for later purposes we also wish to generalise the
solutions by allowing for arbitrary asymptotic values of the six scalar
fields of the STU supergravity.  This can be done by acting with the
remaining six generators of the coset $SL(2,\R)^3/U(1)^3$ within the
global symmetry group.

  One of our reasons for constructing the most general static BPS black hole
solutions explicitly is to study in detail a phenomenon that was first
noticed by Couch and Torrence in 1984 in the case of the extremal
Reissner-Nordstr\"om black hole \cite{coutor}.  This solution, which
is a special case \cite{cvyoII} 
of the static BPS black holes of STU supergravity in
which all four electric charges are set equal and the magnetic charges
are all set to zero, exhibits a conformal inversion symmetry as follows.
Writing the extreme Reissner-Nordstr\"om metric as
\be
ds^2 = -\Big(1+\fft{Q}{r}\Big)^{-2}\, dt^2 + 
         \Big(1+\fft{Q}{r}\Big)^2\, (dr^2+r^2\, d\Omega_2^2)\,,
\ee
where $d\Omega_2^2$ is the metric on the unit 2-sphere, the horizon is
located at $r=0$.  Defined an inverted coordinate $\hat r$ and
a conformally-related metric $d\hat s^2$ by
\be
\hat r = \fft{Q^2}{r}\,,\qquad d\hat s^2 = \fft{Q^2}{r^2}\, ds^2\,,
\ee
one finds that metric $d\hat s^2$ takes exactly the same form as the
original metric $ds^2$, written now in terms of the inverted coordinate:
\be
d\hat s^2 = -\Big(1+\fft{Q}{\hat r}\Big)^{-2}\, dt^2 +
         \Big(1+\fft{Q}{\hat r}\Big)^2\, (d\hat r^2+\hat r^2\, d\Omega_2^2)\,.
\ee
The Couch-Torrence symmetry therefore maps the near-horizon region to the
asymptotic region near infinity, and vice versa.  This symmetry has been
employed in more recent times to related conserved quantities on the
horizon to conserved quantities at infinity.  (See, for example, 
\cite{bizfri,lumureta,godgodpop}.)

Recently, generalisations of the Couch-Torrence symmetry have been studied
for the static black hole solutions of STU supergravity.  In \cite{godgodpop}
it was shown that the set of solutions carrying four electric charges
and no magnetic charges maps into itself under a similar conformal inversion.
The mapping is no longer a symmetry, in that a black hole with charges
$(Q_1,Q_2,Q_3,Q_4)$ maps into a different member of the 4-charge
family, with different charges $(\hat Q_1,\hat Q_2,\hat Q_3,\hat Q_4)$, 
related by $\hat Q_i = Q_i^{-1}\, (Q_1\,Q_2\,Q_3\,Q_4)^{1/2}$ 
\cite{godgodpop}. More recently, we studied the problem for the
general 8-charge static BPS black holes \cite{cvposa1}.  Our focus 
was on the metrics where the asymptotic values of the scalar fields
were taken to vanish.  We established that any such 8-charge black hole
will certainly be related to other members of this family under a conformal
inversion, although it does not seem to be possible any longer to present
an explicit mapping of the charges in the way that could be done for
the 4-charge solutions.  
 
A slightly different approach was developed in \cite{borsduff2}, which
could be applied not only to the static BPS black holes of STU supergravity,
but more generally in Einstein-Maxwell-scalar theories of type $E_7$
\cite{borsduff2}.   An important aspect of this approach is that it is
applied in the context of the general class of static BPS black holes
including the parameters associated with the asymptotic values of the
scalar fields.  We exhibit this description of the conformal inversion
mapping in detail for the STU supergravity case, where it allows
an explicit mapping of the general 14-parameter family of solutions
(comprising 8 charges plus 6 asymptotic scalar values).  At the price
of the conformal inversion always entailing a mapping from one set of
asymptotic scalar values to a different set of values, the approach allows
one to give fully explicit transformations under the inversion.

 The organisation of this paper is as follows.
In section 2, we discuss how the four-dimensional STU supergravity can be 
obtained by Kaluza-Klein reduction from higher dimensions, where the
internal space is a torus.  In particular, it suffices for our purposes
to start from the pure ${\cal N}=2$ non-chiral supergravity in six 
dimensions (which itself can be obtained from a toroidal reduction of
supergravity in ten dimensions), and then perform a reduction on a 2-torus
to four dimensions.  In addition to discussing the resulting formulation
of STU supergravity, known sometimes as the heterotic formulation, 
we also carry out explicitly the dualisation of two of the four gauge fields.  
This results in the formulation known as the U-duality 
basis.  This is the form in which the theory would arise if one began
with the gauged ${\cal N}=8$ supergravity coming from the reduction of
eleven-dimensional supergravity on $S^7$, truncated this to ${\cal N}=2$
supergravity plus three vector supermultiplets, and then took the ungauged 
limit.  We also discuss the STU supergravity in a third duality complex,
which we refer to as the 3+1 formulation, where just one of the field 
strengths of the heterotic formulation is dualised. 
We shall make use of all three of these formulations of STU supergravity,
and the explicit relations between them,
in the remainder of the paper.

In section 3, we consider the general extremal BPS static black hole
solutions of STU supergravity in the U-duality formulation.  
These solutions carry 8 charges in total, comprising 
electric and magnetic charges for each of the four gauge fields.  These
are really equivalent to solutions with just 5 independent charges, 
since the $U(1)^3$ subgroup of the global $SL(2,\R)^3$ symmetry 
group of the STU theory allows 3 of the 8 charges to be transformed
away (while maintaining the vanishing of the asymptotic values of the 
scalar fields at infinity).  In fact the 8-charge solutions were 
constructed in \cite{cvettsey},
in the heterotic formulation, by starting from a 5-charge ``seed solution'' 
and then acting with the $U(1)^3$ transformations.  The analogous process
was employed recently in \cite{cvposa1} to construct the 8-charge solution
in the U-duality formulation of STU supergravity.
The new feature that we implement now in section 3 is to allow also for
arbitrary asymptotic values of the six scalar fields in the 8-charge black
hole solutions.\footnote{Partial results including the scalars were obtained
in \cite{cvettsey}.}  This can be achieved by employing the
action of the remaining six symmetries in the coset $SL(2,\R)^3/U(1)^3$.  

In section 4, we carry out the same steps of generating the general static BPS
black hole solutions in the heterotic formulation.  Again, we first ``fill
out'' the 5-charge seed solution to the full set of 8 charges by
acting with the $U(1)^3$ subgroup of the global symmetry group.  We highlight similarities
and also differences with the description in the U-duality formulation.

In section 5 we describe the consistent truncation of the STU supergravity to
a supersymmetric theory whose bosonic sector comprises the metric, two gauge
fields and two scalar fields (a dilaton and an axion). The theory is  considerably
simpler to work with than the full STU supergravity, and as we show, the
general static BPS black hole solutions, characterised by four charges and
asymptotic values for the two scalar fields, are very much simpler.  

In section 6, we consider the static BPS black holes in the 3+1 formulation,
using the description of the theory in terms of the K\"ahler geometry of
the scalar manifold.  We then use this to investigate the behaviour of
the black hole metrics under an inversion of the radial coordinate, which,
together with a conformal rescaling, maps the horizon to infinity and 
vice versa.  In this description of the conformal inversion, which was
discussed previously in \cite{borsduff2}, one can show how any
given member of the general 14-parameter family of static BPS black holes
(characterised by the 4 electric charges, 4 magnetic charges and 
6 asymptotic scalar values) is mapped by the conformal inversion to another
member of the 14-parameter family.  In this description, unlike one 
considered previously in \cite{godgodpop,cvposa1}, the asymptotic values of the
scalar fields are always different in the original and the conformally-inverted 
metrics.

  Section 7 contains a summary of our conclusions, and also further
discussion.  Some details of the Kaluza-Klein reduction from six to
four dimensions are relegated to appendix A.

\section{The STU Supergravity Theory}

  In this section, we shall provide explicit expressions for the 
bosonic sector of four-dimensional ungauged STU supergravity in the three
different formulations we shall be using in this paper, and the explicit
relations between them.  Specifically, we shall consider
what may be called the heterotic formulation, which is the way the theory
arises if one starts from ten-dimensional
supergravity and toroidally reduces on $T^6$, together with 
appropriate truncations.  The truncated theory has ${\cal N}=2$
supersymmetry, and comprises ${\cal N}=2$ supergravity coupled to three
vector multiplets.  The theory was presented in this form in
\cite{dulira}.   We shall also consider the formulation of STU
supergravity that one would obtain by reducing eleven-dimensional supergravity
on $S^7$, performing an appropriate truncation from ${\cal N}=8$ to 
${\cal N}=2$ supersymmetry, and also turning off the gauge coupling constant
in the four-dimensional theory.  This formulation of STU supergravity is
sometimes referred to as the $U$-duality invariant formulation.  The
two formulations are related in four dimensions by performing an appropriate
dualisation of two of the four gauge fields.  We shall also consider
STU supergravity in an intermediate formulation which we refer to as the
3+1 formulation.  In this case, instead of dualising two field strengths in
the heterotic formulation only one is dualised.  

\subsection{STU supergravity from six dimensions}

  In the heterotic formulation one can in fact conveniently describe the
STU theory by first reducing from ten dimensions to six dimensions on $T^4$
and truncating to pure ${\cal N}=2$ non-chiral supergravity.  The STU
theory is then obtained by reducing this on $T^2$ with no further truncation.
(Except, of course, the usual Kaluza-Klein truncation in which only the
singlets under the $U(1)^2$ isometry of the $T^2$ are retained.)  The bosonic
sector of the non-chiral supergravity in six-dimensional supergravity is
described by the Lagrangian
\bea
{\cal L}_6 =\hat R\, {\hat *\oneone} -\ft12 {\hat *d\hat\phi}\wedge d\hat\phi -
   \ft12 e^{-\sqrt2\, \hat\phi}\, {\hat *\hat H_\3}\wedge \hat H_\3\,,
\label{lag6}
\eea
where $\hat H_\3 = d\hat B$.  The reduction down to four dimensions is
described in detail in appendix \ref{KKredsec}.  After dualising the
2-form potential for the field $H_\3$ to an axion $\chi_1$, the
resulting Lagrangian for the bosonic sector of STU supergravity is given by
(\ref{lagt}):
\bea
{\cal L} &=& R\,{*\oneone} -\ft12 \sum_{i=1,3}({*d\varphi_i}\wedge d\varphi_i
+ e^{2\varphi_i}\, {*d\chi_i}\wedge d\chi_i) 
 -\ft12 {*d\tvp_2}\wedge d\tvp_2 -
  \ft12 e^{2\tvp_2}\, {*d\tchi_2}\wedge d\tchi_2
\nn\\
&& -\ft12 e^{-\varphi_1}\, \Big[
 e^{\tvp_2-\varphi_3}\,{*\tbF}_1\wedge \tbF_1 +
  e^{-\tvp_2+\varphi_3}\,{*\tbF}_2\wedge \tbF_2 \label{lagtt}\\
&&
\qquad\qquad
 + e^{-\tvp_2-\varphi_3}\, {*\bF}^3\wedge\bF^3 +
  e^{\tvp_2+\varphi_3}\, {*\bF}^4\wedge\bF^4 \Big]
+ \chi_1\, (\tF_1\wedge\tF_2 + F^3\wedge F^4)\,,\nn
\eea
where $\tF_1=d\tA_1$, $\tF_2=d\tA_2$, $F^3=dA^3$ and $F^4=dA^4$ are
the ``raw'' field strengths, and
\bea
\tbF_1&=&\tF_1-\tchi_2\, F^3\,,\qquad  \tbF_2=\tF_2 - \chi_3\, F^3\,,\nn\\
\bF^3&=& F^3\,,\qquad \bF^4= F^4 + \chi_3\, \tF_1 +\tchi_2\, \tF_2 -
  \chi_2\, \chi_3\, F^3\,,\label{tildeFF}
\eea
with overbars, 
denote the ``dressed'' field strengths that appear in the kinetic terms in
(\ref{lagtt}).
The gauge fields numbered 1 and 2 are denoted with tildes here; later on, we
shall dualise these fields in order to obtain the STU supergravity theory
in the formalism in which the $SL(2,\R)^3$ symmetry is manifest.

\subsection{Heterotic formulation of STU supergravity}
\label{heteroticsec}

  In this formulation the raw field strengths $(\tF_1,\tF_2,F^3,F^4)$ are
organised into the column vector
\be
\boldsymbol{\cF} \equiv 
\begin{pmatrix} \cF^1 \cr \cF^2 \cr \cF^3 \cr \cF^4 \end{pmatrix} =
\begin{pmatrix} F^3 \cr \tF_1 \cr F^4 \cr \tF_2\end{pmatrix}\,,
\label{cvetvec}
\ee
and the $(\varphi_1,\chi_1)$ dilaton/axion pair are redefined as
\be
\Phi=\ft12\varphi_1\,,\qquad \Psi=\chi_1\,.
\ee
The Lagrangian (\ref{lagtt}) can then be written as
\bea
{\cal L}&=& R\ {*\oneone} +\ft18 \Tr({*dM}\wedge L dM L) 
- 2{*d\Phi}\wedge d\Phi -\ft12 e^{4\Phi}\, {*d\Psi}\wedge d\Psi
\nn\\
&& -\ft12 e^{-2\Phi}\, (LML)_{ij}\,  {*\cF^i}\wedge \cF^j + 
\ft12 \Psi\, L_{ij}\, \cF^i \wedge \cF^j\,.\label{ctlag2}
\eea
where
\be
L=\sigma_1\otimes \mbI_2 =
\begin{pmatrix} 0&\mbI_2\cr \mbI_2 & 0\end{pmatrix}\,.
\label{cvtsL}
\ee
The scalar matrix $M$ can be read off by comparing the kinetic terms for the
gauge fields $\cF^i$ in (\ref{ctlag2}) with the kinetic terms in (\ref{lagtt}).
It is straightforward to see that $M$ can then be written as
\be
M=\begin{pmatrix} \mbG^{-1} & - \mbG^{-1}\, \mbB\cr \mbB \, \mbG^{-1} &
                  \mbG-\mbB\, \mbG^{-1}\, \mbB
\end{pmatrix}
\,,\label{O22Mdef}
\ee
where
\be
\mbG=e^{-\varphi_3}\,
\begin{pmatrix} e^{-\tvp_2} + \tchi_2^2\, e^{\tvp_2} &
               -\tchi_2\, e^{\tvp_2} \cr
              -\tchi_2\, e^{\tvp_2} & e^{\tilde\varphi_2} \end{pmatrix}\,,
\qquad
\mbB= \begin{pmatrix} 0 & -\chi_3 \cr \chi_3 & 0\end{pmatrix}\,.
\label{cGcB}
\ee
One can recognise $\mbG$ as being associated with the internal metric on 
the 2-torus in the Kaluza-Klein reduction (\ref{metred}), (\ie $ds_2^2 =
\mbG_{ij}\, dz^i\, dz^j$ is the metric enclosed in square brackets 
in (\ref{metred}), in the 2-torus directions), and 
$\mbB$ as being associated with the internal component $A_{\0 12}$ of the
2-form potential in (\ref{B2red}).

  The four-dimensional 
Lagrangian (\ref{ctlag2}) is in the form that was obtained in
\cite{cvettsey}. It is invariant under the $O(2,2)$
transformations (the $T$-duality of the 2-torus compactification):
\begin{equation}
M \longrightarrow \Omega M \Omega^T\,,\qquad 
\cA^i  \longrightarrow \Omega^i{}_j\, \cA^j\,, \label{tdual}
\end{equation}
where $g_{\mu\nu}$ and $S$ are inert, and $\cA^i$ are the potentials for the
field strengths $\cF^i$; that is, $\cF^i=d\cA^i$. The transformation matrix 
$\Omega \in O(2,2)$ preserves the $O(2,2)$-invariant matrix $L$:
\begin{equation}
\Omega^T L \Omega = L \ , \ \ \ \ \
L =
\begin{pmatrix}0 & \mathbb{I}_2\cr
\mathbb{I}_2 & 0\end{pmatrix}
\label{mvie}\,.
\end{equation}
$L$ is in fact the metric tensor in the $(2,2)$-signature flat space on which
the $O(2,2)$ transformations $\Omega$ act.  $L$ has components $L_{ij}$
with $L_{13}=L_{31}=L_{24}=L_{42}=0$ and the remainder being zero.  The
inverse metric $L^{-1}$ has components $L^{ij}$, and for these too
$L^{13}=L^{31}=L^{24}=L^{42}=0$ with the remainder being zero.
 Note that one also has 
\begin{equation}
  L \, M\, L = M^{-1}
  \, . \label{Minv}
\end{equation}
Note that $M$ has components $M^{ij}$ and its inverse $M^{-1}$ has components
$(M^{-1})_{ij}= L_{ik} \, L_{j\ell} \, M^{k\ell}$.
  
The equations of motion and Bianchi identities are in addition
invariant under the $SL(2,\R)$ transformations (electromagnetic $S$-duality):
\begin{equation}
S \longrightarrow  {{aS+b}\over{cS+d}}\,,\qquad
{\cal F}^i \to
(c\Psi + d)\,\cF^i  + c \,e^{-2\Phi} \,(ML)_{ij}\,
{*\cF^j}\,,
\label{sdual}
\end{equation}
where $g_{\mu\nu}$ and the scalar matrix $M$ are inert.

   The equations of motion for the electromagnetic fields that follow from
(\ref{ctlag2}) are
\be
d\cG_i=0\,,\qquad\hbox{where}\qquad 
\cG_i\equiv  -e^{-2\Phi}\, (LML)_{ij}\, {*\cF^j} +\Psi\, L_{ij}\, \cF^j\,,
\label{cGdef}
\ee
which implies that the canonical electric charges $\vec\alpha$ will be 
given by\footnote{We have changed the overall sign of the definition of
the dual field strength $\cG_i$ in (\ref{cGdef}), relative to the one
in \cite{cvettsey}.  This is for consistency with our conventions for the
Hodge dualisation of differential forms (which is made explicit in eqn (\ref{Hodgecon}), and our conventions in the rest of the paper.}   
\be
\alpha_i=\fft1{4\pi}\, \int \cG_i \,.\label{vecalphadef}
\ee
The magnetic charges $\vec p$ are given by
\be
p^i=\fft1{4\pi}\, \int \cF^i\,.\label{hetpdef}
\ee
It is convenient for later purposes to define also magnetic charges 
$\vec \beta$ with a lowered $O(2,2)$ index:
\be
\beta_i=L_{ij}\, p^j\,.\label{betap}
\ee
These will be referred to later as the ``canonical'' magnetic 
charges.\footnote{In toroidally compactified 
heterotic string theory the canonical electric charges  ${\vec \alpha}$ are 
quantised and span an even  self-dual lattice, subject to the constraint: 
${\vec \alpha}^TL{\vec\alpha}=-2,0, 2, \cdots$. For BPS-saturated 
configurations one further requires ${\vec \alpha}^TL{\vec\alpha}>0$. 
By S-duality the same conditions are  satisfied  for quantised magnetic 
charges ${\vec \beta}$, along with the the condition that when
${\vec \alpha}\propto {\vec \beta}$ with magnetic and
electric charge vector components being co-prime integers
\cite{SEN1}.}
If the asymptotic values of the scalars are taken to be zero, 
one has $S_{\infty} ={\rm  i}$ and $M_{\infty} =  \mathbb{I}_4$.

\subsection{The U-duality formulation of STU supergravity}
\label{sl2r3sec}

  We arrive at this formulation by dualising the gauge fields 
$\tA_1$ and $\tA_2$ appearing in the Lagrangian (\ref{lagtt}).
To do this, we employ the standard procedure of
introducing dual potentials $A^1$ and $A^2$ as Lagrange multipliers, 
and adding the terms $A^1\wedge d\tF_1 + A^2\wedge d\tF_2$ to the
Lagrangian (\ref{lagt}). Up to total derivatives, this is equivalent
to adding 
\be
{\cal L}_{LM}= F^1\wedge \tF_1 + F^2\wedge \tF_2\,,\label{lmlag}
\ee
where $F^1=dA^1$ and $F^2=dA^2$ are the raw dualised field strengths.  
Varying the total Lagrangian with respect
to $\tF_1$ and $\tF_2$, now treated as independent fields, shows
that $\tF_1$ and $\tF_2$ satisfy
\bea
F^1 + \chi_1\, \tF_2 &=& e^{-\varphi_1+\tvp_2}\, 
    (e^{-\varphi_3}\, {*\tbF}_1 + \chi_3\, e^{\varphi_3}\,
   {*\bF^4})\,,\nn\\
F^2 + \chi_1\, \tF_1 &=& e^{-\varphi_1+\varphi_3}\,
    (e^{-\tvp_2}\, {*\tbF}_2 + \tchi_2\, e^{\tvp_2}\,
   {*\bF^4})\,,\label{dualrels0}
\eea
and hence, using (\ref{tildeFF}),
\bea
F^1 + \chi_1\, \tbF_2 +\chi_1\,\chi_3\, \bF^3&=& 
    e^{-\varphi_1+\tvp_2}\,
    (e^{-\varphi_3}\, {*\tbF}_1 + \chi_3\, e^{\varphi_3}\,
   {*\bF^4})\,,\nn\\
F^2 + \chi_1\, \tbF_1 +\chi_1\, \tchi_2\, \bF^3&=& e^{-\varphi_1+\varphi_3}\,
    (e^{-\tvp_2}\, {*\tbF}_2 + \tchi_2\, e^{\tvp_2}\,
   {*\bF^4})\,,\label{dualrels}
\eea
These two equations, together with their Hodge duals, can be solved 
algebraically for $\tbF_1$ and $\tbF_2$, with the results expressed,
using (\ref{tildeFF}) again, in terms of $(F^1,F^2,F^3,F^4)$ and
$(*F^1,*F^2,*F^3,*F^4)$.  Since the expressions are a little
unwieldy, we shall not present them here.  

   Having solved for $\tbF_1$ and $\tbF_2$ these may be substituted 
back into the Lagrangian (\ref{lagtt}) plus (\ref{lmlag}), to give
the theory in the dualised form, with the gauge fields $(A^1,A^2,A^3,A^4)$.  
After one final step,
in which the $\tvp_2$ and $\tchi_2$ scalars are subjected to the 
discrete involutive $SL(2,\R)$ transformation
\be
\tilde\tau_2 = -\fft1{\tau_2}\,,\qquad\hbox{where}\quad
  \tilde\tau_2=\tchi_2 + i\, e^{-\tvp_2}\,,\qquad
  \tau_2=\chi_2 + i\, e^{-\varphi_2}\,,\label{involution0}
\ee
one finds that the Lagrangian is precisely equal to the one described in
appendix B of \cite{10auth} and appendix A of \cite{cvposa1}.  (The dualisation
to go from the heterotic to the U-duality formulation has also been discussed
in \cite{chowcomp2}.)
The kinetic terms involving the
field strengths given, as in eqn (A.2) of \cite{cvposa1}, by
\be
{\cal L}_F= -\ft12 f^R_{AB}\, {*F}^A\wedge F^B - 
\ft12 f^I_{AB}\, F^A\wedge F^B \,,
\ee
where $f^R_{AB}$ and $f^I_{AB}$ are the real and imaginary parts of the
scalar matrix given in eqn (A.9) of \cite{cvposa1}.   This is written in
the formalism and notation described in Freedman and Van Proeyen \cite{FVP},
in which, defining 
\be
G_A= -f^R_{AB}\, {*F^B} - f^I_{AB}\, F^B\,,\label{Gdef}
\ee
the field equations $dG_A=0$ and Bianchi identities $dF^A=0$ are invariant
under the transformations
\bea
\begin{pmatrix}{\bf F}\cr {\bf G} \end{pmatrix} \longrightarrow
  {\cal S}\, \begin{pmatrix} {\bf F}\cr {\bf G} \end{pmatrix}\,,\qquad
{\cal S}=\begin{pmatrix} A & B\cr C & D\end{pmatrix}\,,\label{FHsp8}
\eea
where the constant $4\times 4$ matrices $A$, $B$, $C$ and $D$ obey 
\be
A^T C - C^T A=0\,,\qquad B^T D - D^T B=0\,,\qquad 
  A^T D - C^T B = \mbI_4\,,\label{ABCD}
\ee
and we have defined ${\bf F}=(F^1,F^2,F^3,F^4)^T$ and 
${\bf G}=(G_1,G_2,G_3,G_4)^T$.
The equations (\ref{ABCD}) 
are precisely the conditions for the matrix ${\cal S}$
to be an element of $Sp(8,\R)$, obeying \cite{FVP}
\be
S^T\Omega S=\Omega\,,\qquad \hbox{where}\qquad \Omega=
\begin{pmatrix} 0&\mbI_4\cr -\mbI_4 &0\end{pmatrix}\,.
\ee
The scalar field Lagrangian 
\be
{\cal L}_{\rm scal} = -\ft12 \sum_{i=1}^3({*d\varphi_i}\wedge d\varphi_i
+ e^{2\varphi_i}\, {*d\chi_i}\wedge d\chi_i)
\ee
is invariant under $SL(2,\R)^3$, which is a 
subgroup of $Sp(8,\R)$, and so in this
formulation the STU supergravity theory has a manifest $SL(2,\R)^3$ 
symmetry, at the level of the equations of motion.  The explicit forms of
the $A$, $B$, $C$ and $D$ matrices corresponding to the $SL(2,\R)^3$ subgroup
of $Sp(8,\R)$ are given in \cite{cvposa1}.  

  The conserved electric and magnetic charges in the U-duality 
formalism are given by
\be
Q_A= \fft1{4\pi}\, \int G_A\,,\qquad P^A=\fft{1}{4\pi}\, \int F^A\,.
\label{PQdef}
\ee
If we define the charge vectors ${\bf P}=(P^1,P^2,P^3,P^4)^T$ and 
${\bf Q}=(Q_1,Q_2,Q_3,Q_4)$ then the action of $SL(2,\R)^3$ on the charges
takes the same form as for the fields and their duals in (\ref{FHsp8}),
namely
\bea
\begin{pmatrix}{\bf P}\cr {\bf Q} \end{pmatrix} \longrightarrow
  {\cal S}\, \begin{pmatrix} {\bf P}\cr {\bf Q} \end{pmatrix}\,,\label{PQsp8}
\eea
where again the $4\times 4$ matrices $A$, $B$, $C$ and $D$ that form ${\cal S}$
are restricted
to the $SL(2,\R)^3$ subgroup of $Sp(8,\R)$, as given in \cite{cvposa1}.

  There is in fact a simpler way to present the action of the $SL(2,\R)^3$
global symmetries on the charges, by introducing the charge tensor
$\gamma_{a a' a''}$, where $a$, $a'$ and $a''$ are doublet indices
of the three $SL(2,\R)$ factors in the symmetry group (see, for example,
\cite{dulira}).  If we make the assignments
\bea
\gamma_{000}&=& -P^1\,,\qquad \gamma_{011}=P^2\,,\qquad 
  \gamma_{101}=P^3\,,\qquad \gamma_{110}= P^4\,,\nn\\
\gamma_{111} &=& -Q_1\,,\qquad \gamma_{100}=Q_2\,,\qquad 
\gamma_{010}= Q_3\,,\qquad \gamma_{001}=Q_4\,,\label{gammatensor}
\eea
then the $SL(2,\R)^3$ transformation of the charges is given by
\be
\gamma_{aa'a''}\longrightarrow (S_1)_a{}^b\, (S_2)_{a'}{}^{b'}\,
   (S_3)_{a''}{}^{b''}\, \gamma_{b b' b''}\,,
\ee
where
\be
S_i = \begin{pmatrix} a_i & b_i\cr c_i & d_i\end{pmatrix}\,,\qquad
a_i\, d_i -b_i\, c_i=1\quad \hbox{for each}\ i\,.
\ee
This gives exactly the same $SL(2,\R)^3$ transformation as in 
appendix A of \cite{cvposa1}.  It furthermore demonstrates that 
$\gamma_{aa'a''}$ is covariant with respect to $SL(2,\R)^3$ transformations.

  One can, of course, conveniently write the $SL(2,\R)^3$ transformation 
(\ref{FHsp8}) in the analogous way, by defining the field strength tensor
\bea
\Phi_{000}&=& -F^1\,,\qquad \Phi_{011}=F^2\,,\qquad
  \Phi_{101}=F^3\,,\qquad \Phi_{110}= F^4\,,\nn\\
\Phi_{111} &=& -G_1\,,\qquad \Phi_{100}=G_2\,,\qquad
\Phi_{010}= G_3\,,\qquad \Phi_{001}=G_4\,,
\eea
which transforms according to
\be
\Phi_{aa'a''}\longrightarrow (S_1)_a{}^b\, (S_2)_{a'}{}^{b'}\,
   (S_3)_{a''}{}^{b''}\, \Phi_{b b' b''}\,,
\ee
Furthermore, if the dilaton/axion pairs of scalar fields $(\varphi_1,\chi_1)$, 
$(\varphi_2,\chi_2)$ and $(\varphi_3,\chi_3)$ are assembled into the
matrices
\be
M_i = \begin{pmatrix} e^{\varphi_i} & -\chi_i\, e^{\varphi_i}\cr
    -\chi_i\, e^{\varphi_i} & e^{-\varphi_i} +\chi_i^2\, e^{\varphi_i}
   \end{pmatrix}\,,\qquad \hbox{for}\qquad i=1,2,3\,,\label{Mmatrices}
\ee
whose components are $(M_1)^{ab}$, $(M_2)^{a'b'}$ and $(M_3)^{a''b''}$
respectively, 
then the scalars transform under the $SL(2,\R)$ factors according to
\be
M_i\longrightarrow M_i' =(S_i^T)^{-1}\, M_i\,S_i^{-1}\,,
\ee
(with the $i$'th scalars transforming only under the $i$'th $SL(2,\R)$ group).  

\subsection{Mapping between the heterotic and the U-duality formulations}

  As we saw in section 2.2, the scalar fields in the heterotic formulation
comprise the dilaton/axion pairs $(\Phi,\Psi)$,
 $(\varphi_2,\chi_2)$ and $(\varphi_3,\chi_3)$.  These
come from the Kaluza-Klein reduction from six dimensions, as described
in appendix A, with $\Phi=\ft12\varphi_1$ and $\Psi=\chi_1$.  The 
pairs $(\varphi_2,\chi_2)$ and $(\varphi_3,\chi_3)$ are packaged into
the $O(2,2)/U(1)^2$ scalar coset matrix $M$ given by (\ref{O22Mdef})
and (\ref{cGcB}), while $(\Phi,\Psi)$ parameterise the $SL(2,\R)/U(1)$
scalar coset.

In the U-duality formulation the scalars comprise the same set
$(\varphi_1,\chi_1)$, $(\varphi_2,\chi_2)$ and $(\varphi_3,\chi_3)$, except
that now the $(\varphi_2,\chi_2)$ pair are subjected to the involution
$\tau_2\rightarrow -1/\tau_2$, where $\tau_2=\chi_2 +i e^{-\varphi_2}$,
as given in (\ref{involution0}).  Thus, in total, the mapping of 
fields from those of the heterotic formulation and to those of
the U-duality formulation is as follows:
\bea
\Phi &\longrightarrow &\ft12\varphi_1\,,\qquad 
\Psi\longrightarrow \chi_1\,,
\qquad \varphi_3\longrightarrow \varphi_3\,,\qquad
\chi_3\longrightarrow \chi_3\,,\nn\\
e^{\tvp_2}&\longrightarrow& 
  (1+\chi_2^2\, e^{2\varphi_2})\, e^{-\varphi_2}\,,
\qquad
  \tchi_2\longrightarrow - \chi_2\, e^{2\varphi_2}\, 
(1+ \chi_2^2\, e^{2\varphi_2})^{-1}\,.\label{hetscalUdual}
\eea

  The relation between the electric and magnetic charges in the two
formulations follows from the relations between the field strengths,
which we discussed earlier.  The dual field strengths $\cG_i$ in the
heterotic formulation, given by (\ref{cGdef}), can be seen, after making
use of the equations (\ref{dualrels}) to express
the fields $\widetilde F_1$ and $\widetilde F_2$ in (\ref{cvetvec}) in terms 
of the dual fields $F^1$ and $F^2$, to be related to the fields $F^A$ and
dual fields $G_A$ of the U-duality formulation by
\be
\cG_1= G_3\Big|_{\tau_2\rightarrow -1/\tau_2}\,,\qquad
\cG_2=-F^1\,,\qquad 
\cG_3= G_4\Big|_{\tau_2\rightarrow -1/\tau_2}\,,\qquad
\cG_4=-F^2\,,\label{cGtoU}
\ee
where in addition to the involution of $(\varphi_2,\chi_2)$ indicated here,
we also make the replacements 
$\Phi\rightarrow \ft12\varphi_1$ and $\Psi\rightarrow \chi_1$,
as given in (\ref{hetscalUdual}).  Similarly, one finds that the 
fields $\cF^i$ of the heterotic formulation are related to $F^A$ and $G_A$ 
by
\be
\cF^1= F^3\,,\qquad \cF^2=G_1\Big|_{\tau_2\rightarrow -1/\tau_2}\,,\qquad
\cF^3= F^4\,,\qquad \cF^4=G_2\Big|_{\tau_2\rightarrow -1/\tau_2}\,.
\label{cFtoU}
\ee
From the definitions (\ref{vecalphadef}) and (\ref{hetpdef}) for the
electric and magnetic charges $\alpha_i$ and $p^i$ in the heterotic 
formulation, and the definitions (\ref{PQdef}) for the electric and
magnetic charges $Q_A$ and $P^A$ in the U-duality formulation we therefore
have
\be
\val=(\alpha_1,\alpha_2,\alpha_3,\alpha_4)=(Q_3,-P^1,Q_4,-P^2)
\label{hetalUdual}
\ee
and 
\be
\vec p=(p^1,p^2,p^3,p^4)= (P^3,Q_1,P^4,Q_2)\,.\label{hetptoUdual}
\ee
It is also useful to record the mapping from the magnetic charges with
lowered index in the heterotic formulation, $\beta_i=L_{ij}\, p^j$,
for which we therefore have
\be
\vbe=(\beta_1,\beta_2,\beta_3,\beta_4)= (P^4,Q_2,P^3,Q_1)\,.
\label{hetbetoUdual}
\ee

\subsection{STU supergravity in the $3+1$ formulation}
\label{3+1sec}

There is a third choice of duality frame for STU supergravity that
is useful for some purposes.  In this frame, which we refer to as the
``$3+1$ formalism,'' the starting point is again the STU supergravity 
Lagrangian
given in eqns (\ref{lagtt}) and (\ref{tildeFF}).  We then dualise just
the $\tF_2$ field.  in this formalism, therefore, the simple
4-charge BPS solution would carry 3 electric charges and 1 magnetic (or vice
versa). 

  To perform the dualisation of $\tF_2$, we add a Lagrange 
multiplier term ${\cal L}_{LM}= F^2\wedge \tF_2$, where $F^2=dA^2$ is the
dualised field, and then solve the algebraic equation of motion obtained
by varying $F^2$.  This allows us to solve for the 
dressed field $\tbF_2=\tF_2 -\chi_3\, F^3$, finding
\be
\tbF_2 = -\fft1{(1+\tchi_2^2\, e^{2\tvp_2})}\, \Big[
   e^{\varphi_1+\tvp_2+\varphi_3}\, ({*F}^2 + \chi_1\, {*\tF}_1) +
             \tchi_2\, e^{2\tvp_2}\, (F^4 + \chi_3\, \tF_1)\Big]\,.
\ee
Substituting back into the original Lagrangian plus the Lagrange multiplier
term, we obtain the bosonic STU supergravity Lagrangian written in
the $3+1$ formulation.  It is convenient
to define the new $(\varphi_2,\chi_2)$ scalars related to
$(\tvp_2,\tchi_2)$ by the involution $\tilde \tau_2=-1/\tau_2$, 
exactly as in (\ref{hetscalUdual}), and
the three ``dressed'' field strengths:
\be
\hat F^2=F^2 + \chi_1\, \tF_1\,,\qquad
\hat F^3=F^3 + \chi_2\, \tF_1\,,\qquad
\hat F^4=F^4 + \chi_3\, \tF_1\,.
\ee
In terms of these, the $3+1$ STU Lagrangian takes the 
form\footnote{A dualisation to obtain the STU theory in the 3+1 formulation
can also be found in \cite{chowcomp1}.}
\bea
{\cal L} &=& R\,{*\oneone} -\ft12 \sum_{i=1}^3({*d\varphi_i}\wedge d\varphi_i
+ e^{2\varphi_i}\, {*d\chi_i}\wedge d\chi_i)  
 -\ft12 e^{-\varphi_1-\varphi_2-\varphi_3}\,{*\tF}_1\wedge \tF_1\nn\\
&& -
\ft12 e^{\varphi_1-\varphi_2-\varphi_3}\,{*\hat F}^2\wedge \hat F^2 -
\ft12 e^{-\varphi_1+\varphi_2-\varphi_3}\,{*\hat F}^3\wedge \hat F^3 -
\ft12 e^{-\varphi_1-\varphi_2+\varphi_3}\,{*\hat F}^4\wedge \hat F^4
\nn\\
&&
+\chi_1\, F^3\wedge F^4 + \chi_2\, F^2\wedge F^4 +
\chi_3\, F^2\wedge F^3 \nn\\
&&+
 (\chi_2\, \chi_3\, F^2 + \chi_1\,\chi_3\, F^3 +
  \chi_1\, \chi_2\, F^4)\wedge \tF_1 +
\chi_1\, \chi_2\, \chi_3\, \tF_1\wedge\tF_1\,.
\label{lag31}
\eea
Note that $\tF_1$, $F^2$, $F^3$ and $F^4$ are the bare field strengths:
\be
\tF_1=d\tA_1\,,\qquad F^2=dA^2\,,\qquad F^3=dA^3\,,\qquad F^4=dA^4\,.
\ee
Note also that the scalar fields here are exactly the same as the ones in
the U-duality formulation. In this $3+1$ formulation there is a 
permutation symmetry among the sets of fields
\be
(\varphi_1,\chi_1,F^2)\,,\qquad (\varphi_2,\chi_2,F^3)\,,\qquad
(\varphi_3,\chi_3,F^4)\,.
\ee

We shall return to a discussion of the 3+1 formulation of STU supergravity
in section \ref{Ksec}.

\section{General Extremal BPS Static Black Holes In The U-Duality Formulation}

   The most general extremal BPS static black hole solution in STU supergravity
will carry 8 independent charges, since each of the four field strengths
can carry an electric charge and a magnetic charge. Since the theory has
an $SL(2,\R)^3$ global symmetry, this can be used in order to map one 
solution
into another solution that is equivalent under the group action.  The
3-parameter compact subgroup $U(1)^3$ leaves the asymptotic values of the
six scalar fields unchanged, and this means that it suffices to consider
an $8-3=5$ parameter ``seed'' solution, with only 5 independent charges, in
order to fill out the full 8-parameter family by acting with $U(1)^3$.  If
ones starts with a 5-charge seed solution in which the dilatonic and 
axionic scalars vanish asymptotically at infinity, then the 8-charge
solutions obtained in this way will all have vanishing asymptotic
scalars.  One can also then choose to fill out the solution set further by then
acting with the remaining 6-parameter coset $SL(2,\R)^3/U(1)^3$ transformations,
thereby giving arbitrary asymptotic values to the six scalar fields.

  In the subsections below, we shall present the general 8-charge
static BPS black hole solutions in the U-duality formulation, both for vanishing values of the
asymptotic scalar fields and for arbitrary values of the asymptotic
scalar fields.

\subsection{The general 8-charge static BPS metric}
\label{Udualitysec}

   The general 8-charge solution for the case of vanishing asymptotic
scalars was constructed in this formulation in \cite{cvposa1}.  The
starting point was the solution with 5 independent charges that was
constructed in \cite{cvettsey}, after translating it into the 
U-duality formulation.  The five independent charges could be taken to 
be\footnote{In \cite{cvposa1} the magnetic charges were actually taken to
be ${\bf P}= (0,0,p,-p)$, but we are making the equally valid choice
in (\ref{5chspec}) here, for consistency with the choice (\ref{5chargehet})
in the heterotic formulation, given the mapping (\ref{hetptoUdual}).}
\be
{\bf Q}=(\bQ_1,\bQ_2,\bQ_3,\bQ_4)\,,\qquad {\bf P}= (\bp,-\bp,0,0)\,.
\label{5chspec}
\ee
(We are placing bars on the charges in the 5-charge seed solution.)
For this solution, the metric is given by \cite{cvettsey} 
\bea
ds^2 &=& -\fft{r^2}{\sqrt{V}}\, dt^2 + \frac{\sqrt{V}}{{r^2}}\, 
(dr^2 + r^2 d\Omega^2)\,, \label{exm}\\
V&=& r^4 + \alpha\, r^3 + \beta\, r^2 + \gamma\, r + \Delta\,,
\eea
where
\bea
\alpha&=& \sum_i \bQ_i\,,\qquad \beta= \sum_{i<j} \bQ_i \bQ_j -\bp^2\,,\qquad
\gamma=\sum_{i<j<k} \bQ_i \bQ_j \bQ_k - \bp^2\, (\bQ_1+\bQ_2)\,,\nn\\
\Delta &=& \prod_i \bQ_i - \ft14 \bp^2\, (\bQ_1+\bQ_2)^2\,.\label{metABCD}
\eea
This was filled out to give the general 8-charge solution by acting with
the $U(1)^3$ subgroup of the $SL(2,\R)^3$ global symmetry of the STU 
supergravity in \cite{cvposa1}.  It was shown that the constants
$\alpha$, $\beta$, $\gamma$ and $\Delta$ are then given by
\bea
\alpha^2&=&\Big(\sum_A Q_A\Big)^2 +\Big(\sum_A P^A\Big)^2\,,\nn\\
\beta&=& \sum_{A<B} (Q_A Q_B + P^A P^B)\,,\nn\\
\alpha\gamma&=&
4\Delta +\ft12\beta^2 -
\ft12\sum_{A<B} \Big ((P^A P^B)^2 + (Q_A Q_B)^2 + P^{AB}\, Q_{AB}\Big)
-3\prod_A P^A -3\prod_A Q_A \,,\nn\\
&&\nn\\
\Delta&=& \prod_A Q_A +\prod_A P^A +\ft12 \sum_{A<B} P^A Q_A P^B Q_B
   -\ft14 \sum_i (P^A)^2\, (Q_A)^2\,,\label{genABCD2}
\eea
where 
$P^{AB}\equiv  P^A P^B +(\prod_C P^C)/(P^A P^B)$ and $Q_{AB}\equiv
Q_A Q_B + (\prod_C Q_C)/(Q_A Q_B)$ (so $P^{12}=P^1 P^2 + P^3 P^4$, etc.).
Stated equivalently, the expressions for $\alpha^2$, $\beta$, 
$\alpha\gamma$ and 
$\Delta$ given in (\ref{genABCD2}) 
are the unique polynomials of degrees 2, 2, 4 and 4 in the charges,
respectively, that are invariant under $U(1)^3$ and that reduce to 
those following from 
(\ref{metABCD}) under the 5-charge specialisation (\ref{5chspec}).
Note that $\Delta$ is actually invariant under the entire $SL(2,\R)^3$
symmetry group; it is the usual quartic charge invariant of STU supergravity.

  The solution described above has 8 independent charges, with the scalar
fields $(\varphi_i,\chi_i)$ going to zero at infinity.  In order to generalise
the solution to include arbitrary values for the six asymptotic scalars,
we need only act with the symmetry generators in the six-dimensional
coset $SL(2,\R)^3/U(1)^3$.  This is most easily implemented by
using the charge-tensor formulation discussed at the end of section
\ref{sl2r3sec}.  We proceed in two stages, the first being to re-express
the quantities $\alpha$, $\beta$, $\gamma$ and $\Delta$ in (\ref{genABCD2})
in terms of the charge tensor $\gamma_{a a' a''}$.  To do this, 
we first introduce the $SL(2,\R)$-invariant antisymmetric
tensors $\ep^{ab}$, $\ep^{a'b'}$ and $\ep^{a'' b''}$, with $\ep^{01}=$, etc.
These obey
\be
(S_1)_a{}^c\, (S_1)_b{}^d\, \ep^{ab}=\ep^{cd}\,,\qquad \hbox{etc.}
\ee
We also note that since $\alpha$, $\beta$ and $\gamma$ are invariant only
under the $U(1)^3$ subgroup but not the full $(SL2,\R)^3$, we may in addition
employ the Kronecker deltas $\delta^{ab}$, $\delta^{a'b'}$ and 
$\delta^{a''b''}$ in their construction, since $\delta$ is invariant under the
$U(1)$ subgroup of $SL(2,\R)$.  It is straightforward to see that
we may now rewrite $\alpha$, $\beta$, $\gamma$ and $\Delta$ in 
(\ref{genABCD2}) as follows:
\bea
\alpha^2 &=&\delta^{ab}\, \delta^{a' b'}\, \delta^{a'' b''}\,
  \gamma_{a a' a''}\, \gamma_{b b' b''} + 2\beta\,,\label{alten}\\
\beta &=& -\ft12(\delta^{ab}\, \ep^{a'b'}\,\ep^{a'' b''} +
\ep^{ab}\, \delta^{a' b'}\, \ep^{a'' b''} +
 \ep^{ab}\, \ep^{a'b'}\, \delta^{a'' b''})\, \gamma_{a a' a''}\, 
      \gamma_{b b' b''}\,,\label{beten}\\
\gamma &=& \fft1{\alpha}\, \Big[\Delta + \ft12 \beta^2 - Y\Big]\,,
\label{gaten}\\
\Delta &=& \ft18 \ep^{ac}\,\ep^{a'b'}\,\ep^{a''b''}\, \ep^{bd}\, \ep^{c'd'}\,
\ep^{c'' d''}\, \gamma_{a a' a''}\, \gamma_{b b' b''}\, 
  \gamma_{c c' c''}\, \gamma_{d d' d''}\,,\label{Delten}
\eea
where
\bea
Y &=&\ft18 \Big(\delta^{ac}\,\ep^{a'b'}\,\ep^{a'' b''}\,
  \delta^{bd}\, \ep^{c'd'}\,\ep^{c''d''} +
   \delta^{a'c'}\,\ep^{ab}\,\ep^{a'' b''}\,
  \delta^{b'd'}\, \ep^{cd}\,\ep^{c''d''} \nn\\
&&\quad +
   \delta^{a''c''}\,\ep^{a'b'}\,\ep^{a b}\,
  \delta^{b''d''}\, \ep^{c'd'}\,\ep^{cd}\Big)\,
\gamma_{a a' a''}\, 
   \gamma_{b b' b''}\, \gamma_{c c' c''}\, \gamma_{d d' d''}\,.
\eea
Note that, as expected, $\Delta$ in (\ref{Delten}) is written without the
use of the Kronecker deltas, since it is invariant under the entire
$SL(2,\R)^3$ symmetry.

  The final step is to introduce the non-zero asymptotic values for
the scalar fields.  As stated above, this could be accomplished by
acting on the expressions (\ref{alten}) -- (\ref{Delten}) with the
coset symmetries $SL(2,\R)^3/U(1)^3$.  In fact a simpler way, with the
added advantage that it will directly express $\alpha$, $\beta$,
$\gamma$ and $\Delta$ in terms of the asymptotic values of the scalars
$(\varphi_i,\chi_i)$, is to note that the only thing that will cause the
expressions to change under the action of $SL(2,\R)^3/U(1)^3$ is the
fact that the Kronecker deltas are used in the construction of 
$\alpha$, $\beta$ and $\gamma$.  If we replace all the Kronecker deltas
in the expressions for $\alpha$, $\beta$ and $\gamma$ 
by the corresponding scalar matrices (\ref{Mmatrices}), evaluated at infinity,
\bea
\delta^{ab}\longrightarrow \overbar M_1^{ab}\,,\qquad
\delta^{a'b'}\longrightarrow \overbar M_2^{a'b'}\,,\qquad
\delta^{a''b''}\longrightarrow \overbar M_3^{a''b''}\,,\label{deltatoM}
\eea
where the bars on the $\overbar M_i$ indicate that the scalars in 
(\ref{Mmatrices}) are evaluated at infinity,
then the expressions for these quantities will be invariant under the full
$SL(2,\R)^3$ symmetry.  Thus we see that we shall now have
\bea
\alpha^2 &=&\overbar M_1^{ab}\, \overbar M_2^{a' b'}\, \overbar M_3^{a'' b''}\,
  \gamma_{a a' a''}\, \gamma_{b b' b''} + 2\beta\,,\nn\\
\beta &=&  -\ft12(\overbar M_1^{ab}\, \ep^{a'b'}\,\ep^{a'' b''} +
\ep^{ab}\, \overbar M_2^{a' b'}\, \ep^{a'' b''} +
 \ep^{ab}\, \ep^{a'b'}\, \overbar M_3^{a'' b''})\, \gamma_{a a' a''}\,
      \gamma_{b b' b''}\,,\nn\\
\gamma&=& \fft1{\alpha}\,\Big[\Delta + \ft12 \beta^2 -
                   Y\Big]\,,\label{albegascal}
\eea
where now
\bea
Y&=&\ft18 \Big(\overbar M_1^{ac}\,\ep^{a'b'}\,\ep^{a'' b''}\,
  \overbar M_1^{bd}\, \ep^{c'd'}\,\ep^{c''d''} +
   \overbar M_2^{a'c'}\,\ep^{ab}\,\ep^{a'' b''}\,
  \overbar M_2^{b'd'}\, \ep^{cd}\,\ep^{c''d''} \nn\\
&&\quad +
   \overbar M_3^{a''c''}\,\ep^{a'b'}\,\ep^{a b}\,
  \overbar M_3^{b''d''}\, \ep^{c'd'}\,\ep^{cd}\Big)\, \gamma_{a a' a''}\,
   \gamma_{b b' b''}\, \gamma_{c c' c''}\, \gamma_{d d' d''}\,.
\eea
$\Delta$ is unchanged, and is still given by (\ref{Delten}).

\subsection{Scalar fields in the U-duality formulation}\label{Uscalarsec}

   The dilatons and axions in the 5-charge static black solutions in the 
heterotic formulation can be read
off from equations (40), (41) and (42) of \cite{cvettsey}\footnote{The 
labelling of the torus coordinates is opposite in \cite{cvettsey}
to the labelling we are using in this paper, with our coordinates 
$(z^1,z^2)$ in appendix A being equal to $(y^2,y^1)$ in \cite{cvettsey}.
This means that the torus metric and 2-form 
components $G_{11}$, $G_{22}$, $G_{12}$ and $B_{12}$ in eqns
(42) in \cite{cvettsey} should be interpreted as our $\mbG_{22}$,
$\mbG_{11}$, $\mbG_{12}$ and $-\mbB_{12}$ respectively (see eqns
(\ref{O22Mdef}) and (\ref{cGcB})).}, together
with eqn (\ref{cGcB}).  Mapping
to the charges of U-duality formulation using (\ref{hetptoUdual}) and
(\ref{hetalUdual}), we have
\bea
e^{2\Phi}&=& \fft{(r+\bQ_1)(r+\bQ_2)}{\sqrt{V}}\,,\qquad 
 \Psi= -\fft{\bp\, (\bQ_1-\bQ_2)}{2(r+\bQ_1)(r+\bQ_2)}\,,\label{dilax10}\\
e^{\tvp_2} &=& \fft{(r+\bQ_2)(r+\bQ_4)}{\sqrt{V}}\,,\qquad
  \tchi_2 = \fft{\bp\,[r+\ft12(\bQ_1+\bQ_2)]}{(r+\bQ_2)(r+\bQ_4)}\,,
\label{dilax20}\\
e^{\varphi_3} &=& \fft{(r+\bQ_1)(r+\bQ_4)}{\sqrt{V}}\,,\qquad
  \chi_3 = -\fft{\bp\,[r+\ft12(\bQ_1+\bQ_2)]}{(r+\bQ_1)(r+\bQ_4)}\,.
\label{dilax30}
\eea
After mapping into the scalar variables of the U-duality formulation, using 
eqn (\ref{hetscalUdual}), we therefore have
\bea
e^{\varphi_1}&=& \fft{(r+\bQ_1)(r+\bQ_2)}{\sqrt{V}}\,,\qquad
 \chi_1= -\fft{\bp\, (\bQ_1-\bQ_2)}{2(r+\bQ_1)(r+\bQ_2)}\,,\label{dilax1}\\
e^{\varphi_2} &=& \fft{(r+\bQ_1)(r+\bQ_3)}{\sqrt{V}}\,,\qquad
  \chi_2 = -\fft{\bp\,[r+\ft12(\bQ_1+\bQ_2)]}{(r+\bQ_1)(r+\bQ_3)}\,,
\label{dilax2}\\
e^{\varphi_3} &=& \fft{(r+\bQ_1)(r+\bQ_4)}{\sqrt{V}}\,,\qquad
  \chi_3 = -\fft{\bp\,[r+\ft12(\bQ_1+\bQ_2)]}{(r+\bQ_1)(r+\bQ_4)}\,.
\label{dilax3}
\eea

    To find the expressions for the scalar fields in the general 8-charge
solution, we follow the same strategy that we used previously for the
metric.  Namely, we take the expressions above for the scalar fields in
the 5-charge seed solution, and then fill these out to 8-charge solutions
by acting with the $U(1)^3\in SL(2,\R)^3$ global symmetry transformations.  
A new feature that arises here is that the scalar fields, unlike the metric,
themselves transform under the global symmetries, and so these transformations
must be included also in the calculation. To be precise, the dilaton/axion
pair $(\varphi_1,\chi_1)$ transforms under $SL(2,\R)_1$ but is inert under
$SL(2,\R)_2$ and $SL(2,\R)_3$.  Analogous statements apply to 
$(\varphi_2,\chi_2)$ and to $(\varphi_3,\chi_3)$.  

   As in our construction of the metric functions for the 8-charge solution,
we find it convenient to first obtain the scalar solutions for the case
where the asymptotic values of the dilatons and axions are all zero. A
very simple replacement at the final stage of the calculation allows 
the introduction of arbitrary values for the asymptotic scalars.

  The process of promoting the 5-charge seed solution to a full
8-charge solution proceeds in the same we that we described in \cite{cvposa1}.
Because we have made a small adjustment in the conventions in this
paper, in order to allow a direct mapping to equivalent the results in
the heterotic formulation, we need first to record the explicit results
for the $U(1)^3$ rotations that yield the 8-charge solution.  
Specifically, it was convenient in this
paper to change the 5-charge configuration in the U-duality formalism from
the one specified in eqn (B.4) of \cite{cvposa1} to the one specified in
eqn (\ref{5chspec}) of this paper.  (That is, ${\bf P}=(p,-p,0,0)$
rather than ${\bf P}=(0,0,p,-p)$.)  This modifies the expressions that
were found in \cite{cvposa1} when solving for the 5 charges and the three
$U(1)$ angles $\theta_i$ in terms of the 8 generic charges $Q_i$ and
$P^i$.  Thus we solve the 8 equations contained in
\be
\bar\gamma_{a a' a''}=(U_1)_a{}^b\, (U_2)_{a'}{}^{b'}\, 
(U_3)_{a''}{}^{b''}\, \gamma_{b b' b''}          \,,\label{repdef}
\ee
where $\bar\gamma_{a a' a''}$ is the charge tensor for
the 5-charge seed configuration in
(\ref{5chspec}); $\gamma_{a a' a''}$ is the charge tensor of the 
generic 8-charge
configuration, and $U_i$ denote the three $U(1)$ matrices with $\theta_1$,
$\theta_2$ and $\theta_3$ as parameters:
\be
U_i= \begin{pmatrix} \cos\theta_i &\sin\theta_i\cr
                    -\sin\theta_i &\cos\theta_i\end{pmatrix}\,.
\ee
(See eqn (\ref{gammatensor}) for a definition of $\gamma_{a a' a''}$.)
Eqn (\ref{repdef}) thus yields expressions for
the three angles $\theta_i$ and the 5 charges of the seed solution,
expressed 
in terms of the 8 arbitrary charges of the generic 8-charge solution.  We
find
\crampest
\bea
\tan \theta_+ &=&
\fft{(P^1+P^2)-(Q_1+Q_2)\tan \theta_1}{(Q_3+Q_4)+(P^3+P^4)\tan\theta_1}\,,
\qquad
\tan\theta_-=
\fft{(P^3-P^4)+(Q_3-Q_4)\tan \theta_1}{(Q_1-Q_2)-(P^1-P^2)\tan\theta_1}
\,,\\
&&\nn\\
\tan2\theta_1 &=& \fft{2(P^3+P^4)(Q_3+Q_4) -2(P^1+P^2)(Q_1+Q_2)}{
 (P^1+P^2+P^3+P^4)(P^1+P^2-P^3-P^4) -(Q_1+Q_2+Q_3+Q_4)(Q_1+Q_2-Q_3-Q_4)}\,,
\label{thetasol}
\nn
\eea
\uncramp
where $\theta_\pm=\theta_2\pm\theta_3$, and 
\bea
\bar p &=& -\bar\gamma_{000}\,,\quad
\overbar Q_1=-\bar\gamma_{111}\,,\quad
\overbar Q_2= \bar\gamma_{100}\,,\quad 
\overbar Q_3= \bar\gamma_{010}\,,\quad
\overbar Q_4= \bar\gamma_{001}\,, \label{5to8charge}
\eea
where we have placed bars on the 5 seed charges $\overbar Q_i$ and $\bar p$.

   To begin, let us consider the dilaton/axion pair $(\varphi_1,\chi_1)$.  
The first step is to promote the right-hand sides of
eqns (\ref{dilax1}) to 8-charge expressions, by using eqns
(\ref{5to8charge}) together with (\ref{thetasol}).  
We shall write the expressions in (\ref{dilax1}) as 
\be
e^{-\varphi_1} = \fft{\sqrt{V(r)}}{D(r)}\, \qquad 
\chi_1= \fft{\delta}{D(r)}\,,
\ee
where
\bea
  D(r) &=& r^2 + \alpha_D\, r + \beta_D\,,\label{Drdef}\\
\alpha_D^2 &=& (\bQ_1+\bQ_2)^2\,,\qquad
\beta_D = \bQ_1\, \bQ_2\,,\qquad 
\delta =\ft12 \bar p\,(\bar Q_1 -\bar Q_2)\,.
\eea

We
first promote these 5-charge expressions to 7-charge expressions, by
acting with the $U(1)_2$ and $U(1)_3$ rotations, while keeping $\theta_1=0$.
From (\ref{thetasol}), this latter requirement means the 7-charge 
restriction implies the $Q_A$ and $P^A$ must obey
\be
   (P^1+P^2)(Q_1+Q_2)  -(P^3+P^4)(Q_3+Q_4)=0\,.\label{7conU}
\ee
This leads straightforwardly to the 7-charge augmentations
\bea
{\alpha_D^2}_{\sst{(7)}} &=& (Q_1+Q_2)^2+(P^3+P^4)^2\,,\qquad
{\beta_D}_{\sst{(7)}}= Q_1\, Q_2 + P^3\, P^4\,,\nn\\
\delta_{\sst{(7)}} &=& \ft14[(P^1-P^2)(Q_1-Q_2) - (P^3-P^4)(Q_3-Q_4)]\,.
\label{7chfns}
\eea

  The augmentation from 7 to 8 charges, relaxing the constraint (\ref{7conU}),
is achieved by further rotating the quantities in (\ref{7chfns}) under the
remaining $U(1)_1$ transformation (now with $\theta_2$ and $\theta_3$ set to
zero).  It is helpful at this point to establish a general result for the
$U(1)_1$ rotation of a general 2-index symmetric tensor $W_{ab}$; this
obeys the transformation
\bea
W_{ab}\longrightarrow \widetilde W_{ab}=
   (S_1)_a{}^c\, (S_1)_b{}^d\, W_{cd}\,,\label{Wtrans}
\eea
where $S_1$ is the $U(1)_1\in SL(2,\R)_1$ matrix
\be
S_1=\begin{pmatrix} \cos\theta_1 &\sin\theta_1\cr
                    -\sin\theta_1& \cos\theta_1\end{pmatrix}\,.
\ee
If we define the $U(1)_2\times U(1)_3$ -invariant symmetric tensor
\be
Z_{ab}= (\ep^{a'b'}\, \ep^{a'' b''}-\delta^{a'b'}\, \delta^{a'' b''})\,
    \gamma_{a a' a''}\, \gamma_{b b' b''}\,,\label{Zabdef}
\ee
where the charge tensor $\gamma_{a a' a''}$ is defined in 
(\ref{gammatensor}), 
then the expression for $\tan2\theta_1$ in eqn (\ref{thetasol}) can be written
in the compact form
\be
\tan2\theta_1 = \fft{2 Z_{01}}{Z_{00}-Z_{11}}\,.\label{thetaZ}
\ee
From this we find
\be
\cos2\theta_1=\fft{Z_{00}-Z_{11}}{\Xi}\,,\qquad 
\sin2\theta_1=\fft{2 Z_{01}}{\Xi}\,,\label{cossin2th1}
\ee
where $\Xi$ is the $U(1)^3$-invariant quantity defined by
\be
\Xi^2= Z_{ab}\, Z_{cd}\, (\delta^{ac}\, \delta^{bd} -\ep^{ac}\, \ep^{bd})\,.
\label{XiUdef}
\ee
It then follows straightforwardly from (\ref{Wtrans}) that
\bea
\widetilde W_{00} &=& \ft12\Big[\delta^{ab}\, W_{ab} +\fft{G(W)}{\Xi}\Big]
\,,\qquad
\widetilde W_{11}= \ft12\Big[\delta^{ab}\, W_{ab} -\fft{G(W)}{\Xi}\Big]\,,\nn\\
\widetilde W_{01} &=& \fft1{\Xi}\, Z_{ab}\, W_{cd}\, \delta^{ac}\, \ep^{bd}\,,
\label{Wtrans2}
\eea
where
\be
G(W)\equiv W_{ab}\, Z_{cd}\, 
(\delta^{ac}\, \delta^{bd}-\ep^{ac}\, \ep^{bd})\,.\label{GWdef}
\ee
(Note from (\ref{XiUdef}) that $G(Z)=\Xi^2$.)

 To proceed with the augmentation of the 7-charge expressions 
(\ref{7chfns}), we note that if we define the two $U(1)_2\times U(1)_3$ 
-invariant tensors 
\be
X_{ab}=\ep^{a' b'}\, \ep^{a'' b''}\, \gamma_{a a' a''}\, 
   \gamma_{b b' b''}\,,\qquad
Y_{ab}=\delta^{a' b'}\, \delta^{a'' b''}\, \gamma_{a a' a''}\,
   \gamma_{b b' b''}\,,
\ee
where the charge tensor $\gamma_{a a' a''}$ is defined in eqn
(\ref{gammatensor}),
then the 7-charge expressions in (\ref{7chfns}) can be written as
\be
{\alpha_D^2}_{\sst{(7)}}= -\widetilde Z_{11}\,,\qquad 
{\beta_D}_{\sst{(7)}}= -\ft12 \widetilde X_{11}\,,\qquad
\delta_{\sst{(7)}}=\ft14(\widetilde X_{01} + \widetilde Y_{01})\,.
\ee
Using the $U(1)_1$ transformations (\ref{Wtrans2}),\footnote{The transformation
(\ref{Wtrans2}) is viewed here as the {\it inverse}, taking us from the
general 8-charge (untilded) configuration to the 7-charge (tilded) 
configuration.}
 we therefore obtain
the 8-charge expressions
\bea
\alpha_D^2 &=& -\ft12(\delta^{ab}\, Z_{ab} - \Xi)\,,\nn\\
\beta_D &=& -\ft14\Big[\delta^{ab}\, X_{ab} - \fft{G(X)}{\Xi}\Big]\,,\nn\\
\delta &=& -\fft1{4\Xi}\,(X_{ab}+Y_{ab})\,Z_{cd}\, 
        \delta^{ac}\, \ep^{bd}\,,\label{delalbe}
\eea
and $G(X)$ is given by substituting $W_{ab}=X_{ab}$ in (\ref{GWdef}).
These expressions are all manifestly invariant under $U(1)^3$.  It is
useful to note that $\alpha_D$ can be written in terms of $\alpha$ 
(given in (\ref{alten})) and $\Xi$ as
\be
\alpha_D = \fft{\alpha^2+\Xi}{2\alpha}\,.\label{alDal}
\ee
It is also worth noting that if we define the $U(1)^3$ invariants 
\bea
Z_1&=& -\delta^{ab}\, \ep^{a'b'}\,\ep^{a'' b''}\, \gamma_{a a' a''}\,
      \gamma_{b b' b''} \,,\nn\\
Z_2 &=& -\ep^{ab}\, \delta^{a' b'}\, \ep^{a'' b''}\, \gamma_{a a' a''}\,
      \gamma_{b b' b''}\,,\nn\\
Z_3 &=& -\ep^{ab}\, \ep^{a'b'}\, \delta^{a'' b''}\, \gamma_{a a' a''}\,
      \gamma_{b b' b''}\,,\nn\\
Z_0&=& \delta^{ab}\, \delta^{a'b'}\,\delta^{a'' b''}\, \gamma_{a a' a''}\,
      \gamma_{b b' b''} \,,\label{Zdefs}
\eea
then $\Xi^2$, defined in (\ref{XiUdef}), can be written in the 
factorised form
\be
\Xi^2 =(Z_0+Z_1+Z_2+Z_3)(Z_0+Z_1-Z_2-Z_3)\,.\label{Xisqfactored}
\ee

  It remains to implement the $U(1)^3$ transformations that augment the 
5 charges to 8 charges on the scalar fields.  Since we are looking
specifically at $\varphi_1$ and $\chi_1$, which transform only under 
$U(1)_1$, this means that we just have to make the replacement
\be
\chi_1 +i\, e^{-\varphi_1}\longrightarrow 
  \fft{(\chi_1 +i\, e^{-\varphi_1})\, \cos\theta_1 + \sin\theta_1}{
    \cos\theta_1 - (\chi_1 +i\, e^{-\varphi_1})\,\sin\theta_1}\,.
\ee
In other words, the scalars $\varphi_1$ and $\chi_1$ will be given by
\be
e^{\varphi_1} = \fft{{\cal D}\, D(r)}{\sqrt{V(r)}}\,,\qquad
\chi_1 = \fft{ {\cal C} }{ {\cal D} }\,,\label{phi1chi1trans}
\ee
where
\bea
{\cal C}&=& \fft{\delta}{D(r)}\, \cos2\theta_1 -
         \fft12\Big[\fft{V(r)+\delta^2}{D(r)^2} -1\Big]\, \sin2\theta_1\,,\nn\\
{\cal D} &=& \fft12 \Big[ 1 +\fft{V(r)+\delta^2}{D(r)^2}\Big] -
  \fft{\delta}{D(r)}\, \sin2\theta_1 +
   \fft12\Big[1-\fft{V(r)+\delta^2}{D(r)^2}\Big]\, \cos2\theta_1\,,
\label{cCcDexp}
\eea
where $V(r)$ in (\ref{exm}) is calculated using the 8-charge expressions
(\ref{alten})--(\ref{Delten}), the quantities $\delta$, $\alpha_D$ and 
$\beta_D$ appearing in $D(r)$ are given by the 8-charge expressions  
in (\ref{delalbe}), and $\cos2\theta_1$
and $\sin2\theta_1$ are given by (\ref{cossin2th1}).  

  The expressions for $\varphi_1$ and $\chi_1$ are in fact considerably
simpler than is immediately evident from (\ref{phi1chi1trans}) and 
(\ref{cCcDexp}).  This is because the quadratic function $D(r)$ is actually
a divisor of the quartic function $V(r)+\delta^2$ that appears in (\ref{cCcDexp}).  In fact we can show that
\be
V(r) +\delta^2 = D(r)\, \wtd D(r)\,,\qquad \wtd D(r)=r^2 + \td\alpha_D \, r+
\td \beta_D\,,
\ee
where the coefficients in the quadratic function $\wtd D(r)$ are given by
\be
\td\alpha_D^2 = -\ft12(\delta^{ab}\, Z_{ab} + \Xi)\,,\qquad 
\td\beta_D= -\ft14\Big[ \delta^{ab}\, X_{ab} + \fft{G(X)}{\Xi}\Big]\,.
\label{tdalDtdbetaD}
\ee
(Note that $\td\alpha_D$ and $\td\beta_D$ are closely related
to the coefficients $\alpha_D$ and $\beta_D$ in the 
function $D(r)=r^2+ \alpha_D\, r
+\beta_D$ given in (\ref{delalbe}).)  Thus we see that $\varphi_1$ and $\chi_1$
are given by
\be
e^{\varphi_1} = \fft{r^2 + b_1\, r + b_0}{\sqrt{V(r)}}\,,\qquad
\chi_1 = \fft{c_1\, r + c_0}{r^2+ b_1\, r + b_0}\,,\label{phi1chi1res}
\ee
where
\bea
b_1&=& \fft{\alpha_D + \td\alpha_D}{2} + 
 \fft{(\alpha_D-\td\alpha_D)(Z_{00}-Z_{11})}{2\Xi}\,,\nn\\
b_0&=& \fft{\beta_D+\td\beta_D}{2} -\fft{2\delta\, Z_{01}}{\Xi} +
     \fft{(\beta_D-\td\beta_D)(Z_{00}-Z_{11})}{2\Xi}\,,\nn\\
c_1&=& \fft{(\alpha_D-\td\alpha_D)\,Z_{01}}{\Xi}\,,\qquad
c_0= \fft{\delta\, (Z_{00}-Z_{11})+ (\beta_D-\td\beta_D)\,Z_{01}}{\Xi}\,,
\label{bcdef}
\eea
Using our previous expressions for the various quantities appearing in
(\ref{bcdef}), we find that in terms of the eight charges of the general
static BPS black holes,
\bea
b_1&=& \fft1{\alpha}\, \Big[(P^3+P^4)\, \sum_A P^A+
                     (Q_1+Q_2)\,\sum_A Q_A\Big]\,,\nn\\
b_0&=& P^3\, P^4 + Q_1\, Q_2\,,\nn\\
c_1&=& \fft1{\alpha}\, \Big[ (P^1+P^2)(Q_1+Q_2)  - (P^3+P^4)(Q_3+Q_4)\Big]
\,,\nn\\
c_0&=& \ft12(P^1\, Q_1 + P^2\, Q_2 -P^3\, Q_3 -P^4\, Q_4)\,,
\label{bcres}
\eea
where as usual $\alpha=\sqrt{(\sum_A P^A)^2 + (\sum_A Q_A)^2}$.

  Finally, it remains to introduce non-vanishing asymptotic values for the
scalar fields.  As before when we discussed the metric, 
this is accomplished by making $SL(2,\R)^3/U(1)^3$
coset transformations on the charges, and now also on the scalar fields.  
Thus we introduce an asymptotic  coset vielbein $\overbar\cV_i$ for each 
of the $SL(2,\R)_i/U(1)_i$ coset factors, with
\be
\overbar\cV_i=
  \begin{pmatrix}
     e^{\ft12\bar\varphi_i} & -\bar\chi_i\, e^{\ft12\bar\varphi_i}\cr
           0& e^{-\ft12 \bar\varphi_i}\end{pmatrix}\,,
\ee
where $(\bar\varphi_i,\bar\chi_i)$ denotes the asymptotic values of
the scalar fields. The corresponding transformation of the charge tensor
is therefore
\be
\gamma_{a a' a''}\longrightarrow \overbar\cV^b_{1\, a}\, 
               \overbar\cV^{b'}_{2\, a'}\,
           \overbar\cV^{b''}_{3\, a''}\,\gamma_{b b' b''}\,.\label{gammatrans}
\ee
The effect of this is that all $U(1)^3$ invariants, such as 
$\alpha$, $\beta$, $\gamma$, $\alpha_D$, $\beta_D$, $\delta$, $\td\alpha_D$,
$\td\beta_D$  will 
receive modification, namely that each Kronecker delta in the expressions
(\ref{alten})--(\ref{gaten}), (\ref{Zabdef}), (\ref{Zdefs}), (\ref{bcdef}),
etc., will be replaced
by the corresponding asymptotic scalar matrix 
\be
\overbar M_i= \overbar\cV_i^T\, \overbar\cV_i\,,
\ee
as in eqn (\ref{deltatoM}).  Additionally, the components of the tensor
$Z_{ab}$ appearing in the expressions (\ref{bcdef}) will be transformed
by making the replacement
\be
Z_{ab}\longrightarrow \overbar\cV_1^c{}_a\, \overbar\cV_1^d{}_b\, Z_{cd}\,.
\ee
As already observed, the quantity $\Delta$ 
receives
no modification because its construction in (\ref{Delten}) uses only
the $SL(2,\R)$-invariant epsilon tensors, and no Kronecker deltas.

It is worth recording that even when the asymptotic scalars are non-zero,
the quantity $\Xi$, defined now by (\ref{XiUdef}) with all Kronecker
deltas replaced by the asymptotic scalar $M$ matrices as in (\ref{deltatoM}),
is again simply factorisable, as
\be
\Xi = \alpha\, \tilde\alpha\,,
\ee
where $\alpha$ is given in (\ref{albegascal}) and $\tilde\alpha$ is
the closely related quantity given by
\bea
\tilde\alpha^2 &=&
\Big(\overbar M_1^{ab}\, \overbar M_2^{a' b'}\, \overbar M_3^{a'' b''}\,
  \gamma_{a a' a''}\, \gamma_{b b' b''} 
-\overbar M_1^{ab}\, \ep^{a'b'}\,\ep^{a'' b''} \nn\\
&&\ \ +
\ep^{ab}\, \overbar M_2^{a' b'}\, \ep^{a'' b''} +
 \ep^{ab}\, \ep^{a'b'}\, \overbar M_3^{a'' b''}\Big)\, \gamma_{a a' a''}\,
      \gamma_{b b' b''}\,.
\eea
In other words $\Xi^2$, which is a quartic multinomial in the charges 
with coefficients that are multinomial in the $\bar\chi_i$ and 
$e^{\bar\varphi_i}$ asymptotic scalar values, factorises as the
product of the two quadratic multinomials $\alpha^2$ and $\tilde\alpha^2$.

It is also useful to note that the quantities $\alpha_D$ and $\tilde\alpha_D$,
given by (\ref{delalbe}) and (\ref{tdalDtdbetaD}) after the
replacements (\ref{deltatoM}), are related to $\alpha$ and $\tilde\alpha$ by
\be
\alpha_D+\tilde\alpha_D=\alpha\,,\qquad \alpha_D-\tilde\alpha_D= 
\tilde\alpha\,.
\ee
Thus after introducing the asymptotic scalar values 
the coefficients $b_1$ and $c_1$ in eqns (\ref{bcdef}) are particularly simple,
and are given by
\be
b_1= \fft1{2\alpha}\, \Big[ \alpha^2 + \wtd Z_{00}-\wtd Z_{11}\Big]\,,\qquad
c_1= \fft1{\alpha}\, \wtd Z_{01}\,,
\ee
where $\wtd Z_{ab}= \overbar\cV_1^c{}_a\, \overbar\cV_1^d{}_b\, Z_{cd}$
and all Kronecker deltas involved in the construction of $\alpha$ and
$Z_{ab}$ are as usual 
replaced by $M$ matrices according to (\ref{deltatoM}).

 The action of the coset transformations 
on the scalar fields themselves will be given by sending
\be
M_i\longrightarrow \overbar\cV_i^T\, M_i\, \overbar\cV_i\,,
\qquad \hbox{for}\quad i=1,2,3\,,
\ee
where the $M_i$ scalar matrices are given in (\ref{Mmatrices}).  These
transformations amount to
\be
e^{-\varphi_i}\longrightarrow e^{-\varphi_i-\bar\varphi_i}\,,\qquad
\chi_i\longrightarrow \bar\chi_i + \chi_i\, e^{-\bar\varphi_i}\,.
\label{phichishift}
\ee

  The $U(1)$ rotation angle $\theta_1$ will be modified by
the coset transformations, as dictated by substituting the 
$\gamma_{a a' a''}$ transformations (\ref{gammatrans}) into the
expression (\ref{Zabdef}) for $Z_{ab}$, and then substituting these
transformed $Z_{ab}$ components into (\ref{thetaZ}).  However, since
we have already re-expressed the expressions in (\ref{phi1chi1trans}) 
and (\ref{cCcDexp}) for $\varphi_1$ and $\chi_1$ in the simpler forms
given by (\ref{phi1chi1res}) and (\ref{bcdef}), we no longer need
to implement the coset transformation on $\theta_1$ explicitly.

  The procedure we have described above provides specifically the 
expressions for the dilaton/axion pair $(\varphi_1,\chi_1)$ in the 
general 8-charge static BPS black hole solutions.  One could repeat the
discussion, starting from the same 5-charge seed solution, to 
augment the expressions (\ref{dilax2}) and (\ref{dilax3}) for the
$(\varphi_2,\chi_2)$ and $(\varphi_3,\chi_3 )$ dilaton/axion pairs.  In
fact a simpler way of arriving at the same result is to exploit the
triality symmetry of the U-duality formulation of STU supergravity.  The
precise statement of this triality can be seen from the definition of
the $SL(2,\R)^3$ charge tensor $\gamma_{a a' a''}$ in (\ref{gammatensor}).
The first $SL(2,\R)$ index, $a$, is associated with the $(\varphi_1,\chi_1)$
pair, and the charges $(P^2,Q_2)$; the $a'$ index with $(\varphi_2,\chi_2)$
and the charges $(P^3,Q_3)$, and the $a''$ index with $(\varphi_3,\chi_3)$ 
and the charges $(P^4,Q_4)$.  Thus from the expressions we have 
obtained for $(\varphi_1,\chi_1)$, we just have to permute the labellings 
according to this triality correspondence, in order to obtain
the expressions for $(\varphi_2,\chi_2)$ and $(\varphi_3,\chi_3)$.
        
 Acting with the ${\mathbb Z}_2\in\hbox{triality}$ symmetry 
\be
(\varphi_1,\chi_1; P^2,Q_2) \longleftrightarrow (\varphi_2,\chi_2; P^3,Q_3)
\label{swap12}
\ee
on the results obtained above for the the $\varphi_1$ and $\chi_1$
scalars will give the expressions for $\varphi_2$ and $\chi_2$.
Acting instead with the ${\mathbb Z}_2\in\hbox{triality}$
symmetry
\be
(\varphi_1,\chi_1; P^2,Q_2) \longleftrightarrow (\varphi_3,\chi_3; P^4,Q_4)
\label{swap13}
\ee
will give the expressions for $\varphi_3$ and $\chi_3$.  

   If we specialise for simplicity to the case with vanishing asymptotic values
for the scalar fields, acting with the triality transformations (\ref{swap12})
or (\ref{swap13}) on the constants in (\ref{bcres}) will map 
(\ref{phi1chi1res}) into expressions for $\varphi_2$ and $\chi_2$, or
$\varphi_3$ and $\chi_3$, respectively.

As a check on the calculations in this section, we can take the
expressions of the $\varphi_i$ and $\chi_i$ scalars and make the 5-charge specialisation
given by eqn (\ref{5chspec}).   Doing this, it is straightforward to see that
we do indeed recover the expressions given in eqns (\ref{dilax1}),
(\ref{dilax2}) and (\ref{dilax3}).  Expressed conversely, this shows that if we were 
instead to act with the U-duality transformations in order to elevate
the 5-charge expressions for $(\varphi_2,\chi_2)$ and $(\varphi_3,\chi_3)$
to general 8-charge expressions, we would indeed obtain the results that
follow by making the triality transformations on the
general 8-charge expressions for $(\varphi_1,\chi_1)$ 
that we have constructed.

\subsection{Gauge fields for the 8-charge BPS black holes}

 Having obtained the expressions for the metric and the scalar fields, 
the form of the gauge fields in the general 8-charge static BPS black 
hole solutions follow straightforwardly from the gauge fields equations of
motion. 

 In the U-duality formulation, we find 
\be
F^A= P^A\, \sin\theta\, d\theta\wedge d\varphi +
\fft1{\sqrt{V}}\, \big( (f^{R})^{-1}\big)^{AB}\,
(Q_B + f^I_{BC}\, P^C)\, dt\wedge dr\,,
\ee
which is consistent with the definitions of the electric and magnetic
charges in (\ref{PQdef}).

Note that our conventions for
Hodge dualisation are such that in the metric (\ref{exm}) we have
\be
*(\sin\theta\, d\theta\wedge d\varphi)= \fft1{\sqrt{V}}\, dt\wedge dr\,,
\qquad 
*(dt\wedge dr) = - \sqrt{V}\, \sin\theta\, d\theta\wedge d\varphi\,.
\label{Hodgecon}
\ee

\section{General 8-Charge Static BPS Black Holes In The Heterotic Formulation}

\subsection{8-charge static BPS black hole metric}
\label{hetsec}

   The starting point for writing the general 8-charge static BPS 
black holes with arbitrary asymptotic values for the scalar fields is
again the metric (\ref{exm}).  The charges in the seed solution with
five independent charges and with 
vanishing asymptotic scalars take the form 
\be
\vec \alpha_0=(q_1,-{\bar q} ,q_3,{\bar q})\,,\qquad
\vec p_0=(0,p^2,0,p^4)\label{5chargehet}
\ee
in the heterotic basis \cite{cvettsey}.  (The superscripts 2 and 4 
refer to the indexed labelling of charges $p^i$.  See section 2.2 for the 
notation.) The quantity ${\bar q}$ parameterises the introduction of
the fifth, independent, charge)  The coefficients $\alpha$, $\beta$,
$\gamma$ and $\Delta$ appearing in the metric function $V(r)$ are given 
by \cite{cvettsey}
\begin{eqnarray}
\alpha &=& q_1+ q_3 + p^2 + p^4\,, \nn\\
\beta &=& q_1\, q_3 -{\bar q}^2  +p^2p^4 + (q_1 +q_3)(p^2+p^4)\,, \nn\\
\gamma &=& (q_1 q_3 - {\bar q}^2) (p^2+p^4)  +p^2p^4\,(q_1 +q_3)\,, \nn\\
 \Delta &=& (q_1q_3) (p^2p^4) -\tfrac{1}{4}{\bar q}^2\,(p^2+p^4)^2\,.
\label{5chargealbegaDe} 
\end{eqnarray}
(Here ${\bar q}^2$ means just the square of ${\bar q}$.)

  The general solutions with 8 charges and non-vanishing asymptotic scalars
can be filled out by acting with the global $O(2,2)\times SL(2,\R)$ symmetry
of the heterotic formulation, where $O(2,2)$ is the T-duality symmetry
from the 2-torus and $SL(2,\R)$ is the electric/magnetic S-duality symmetry.
We proceed in three stages; first, acting with the $U(1)^2$ subgroup
of $O(2,2)$ to augment the 5-charge solution to 7 charges; then
acting with the $U(1)$ subgroup of $SL(2,\R)$ to augment further to
8 charges; and finally, acting with the remaining $O(2,2)/U(1)^2$ and 
$SL(2,\R)/U(1)$ cosets in order to introduce the non-vanishing
asymptotic scalars.

\bigskip

\noindent{\underline{Acting with $U(1)^2\in O(2,2)$}:} 
\medskip

The 5-charge starting point can be filled out to 7 independent charges by
acting with the $U(1)^2$ compact subgroup of
$O(2,2)$.  Since $\alpha$, $\beta$ and $\gamma$ should be invariant
under $U(1)^2$ but not under the remaining $O(2,2)/U(1)^2$ coset
action (that is, when the asymptotic scalars vanish), the 2 additional
charges can be introduced by writing $\alpha$, 
$\beta$ and $\gamma$ as $U(1)^2$-invariant expressions
in $\vec \alpha$ and $\vec p$ that reduce to (\ref{5chargealbegaDe}) under
the 5-charge specialisation (\ref{5chargehet}).  In fact, rather
than using the magnetic charges $\vec p$ with components $p^i$ we
shall instead find it convenient to lower the index and work with the
magnetic charges $\vec\beta$, which have the components 
$\beta_i=L_{ij}\, p^j$, as defined in (\ref{betap}).  The available 
``building blocks'' are the matrix $L$ defined in (\ref{mvie}), which is
fully $O(2,2)$ invariant, and the $4\times 4$ identity matrix $\mbI_4$,
which is invariant only under the $U(1)^2$ subgroup.\footnote{Actually
things are a bit more complicated.  There are two further matrices
that are invariant under the $U(1)^2$ subgroup, 
namely $K_1$ and $K_2$ given by
\be
K_1 = \begin{pmatrix} i\sigma_2 &0\cr0&i\sigma_2\end{pmatrix}\,,\qquad
K_2=\begin{pmatrix}0&i\sigma_2\cr i\sigma_2 & 0\end{pmatrix}\,,\label{K1K2}
\ee
where $\sigma_2$ is the usual Pauli matrix.  Note that
$K_1$ and $K_2$ are antisymmetric.  It turns out that by using 
$K=K_1+K_2$, one can write the quantity 
$\Sigma^2$ that appears later in (\ref{Sigsq}) as a manifest 
perfect square. It then gets dressed up
with asymptotic scalars once these are turned on.  See later in
this section for a discussion of this point.\label{Kfoot}} 
   We can also
construct the more general expression for $\Delta$ along similar lines.
Since $\Delta$ is in fact invariant under the full $O(2,2)$ symmetry its
expression will not involve the use of $\mbI_4$.

   Defining the matrices 
\be
\nu_\pm= I_4 \pm L^{-1}\,,\label{nupm}
\ee
one can straightforwardly establish that $\alpha$, $\beta$, $\gamma$ and
$\Delta$ may be written for the 7-charge solutions 
in $U(1)^2$-invariant terms as follows:
\begin{eqnarray}
\alpha^2& =& \val^T \nu_+ \,\val + \vbe^T \nu_+\,\vbe 
   +2 \overbar\Sigma \,, \label{alpha0}\\
2\beta&=&  \val^T L^{-1}\, \val + \vbe^T L^{-1}\, \vbe +
  2\overbar\Sigma \,, \label{beta0}\\
2\alpha \gamma&=&(\val^T L^{-1}\,\val + \vbe^T L^{-1}\,\vbe)\overbar\Sigma 
  \nn\\
&& + (\val^T L^{-1}\,\val)(\vbe^T \nu_+\,\vbe) +
    (\vbe^T L^{-1}\,\vbe)(\val^T \nu_+\,\val)\,,\label{gamma0}\\
4\Delta  &=&(\val^T L^{-1}\,\val)(\vbe^T L^{-1}\,\vbe)- 
  (\val^T L^{-1}\, \vbe)^2\,,\label{Delta0}
\end{eqnarray}
where
\be
\overbar\Sigma^2 = (\val^T\nu_+\, \val)(\vbe^T \nu_+\,\vbe)\,.\label{barSigsq}
\ee
These expressions reduce to those in (\ref{5chargealbegaDe}) under the
5-charge specialisation (\ref{5chargehet}).
Note also that in the original 5-charge configuration (\ref{5chargehet}), the
charges satisfy the constraint 
\be
\val^T \nu_+\,\vbe=0\,,\label{7con}
\ee
and that this continues to be true after filling out to 7 charges, as we have
done by acting with  $U(1)^2$.  

\bigskip

\noindent{\underline{Acting with $U(1)\in SL(2,\R)$}:}
\medskip

   The constraint (\ref{7con}) is removed once the 8th and final charge is
introduced.  This is achieved by acting with the $U(1)$ subgroup of the
remaining $SL(2,\R)$ S-duality symmetry.  Its action on $\val$ and
$\vbe$ is
\begin{equation}
\val\to \val \,\cos\psi - \vbe\, \sin\psi \,, \qquad 
\vbe\to \val\, \sin\psi  + \vbe\, \cos\psi\,, \label{U1}
\end{equation}
where
\begin{equation}
\tan 2\psi= -
\frac{2(\val^T \nu_{+}\, \vbe)}{(\val^T \nu_{+}\, \val)-
 (\vbe^T \nu_{+}\,\vbe)}\,.\label{psi}
\end{equation}
Thus we have 
\be
\cos2\psi= \fft{1}{\Xi}\, (\val^T\nu_+\,\val -\vbe^T\nu_+\,\vbe)\,,
\qquad
\sin2\psi= - \fft{2}{\Xi}\, \val^T\nu_+\,\vbe\,,\label{cossinpsi}
\ee
where
\bea
\Xi^2 &=& (\val^T \nu_+\,\val +\vbe^T \nu_+\,\vbe)^2 -4\Sigma^2\,,
          \label{hetXidef}\\
\Sigma^2 &=& (\val^T\nu_+\, \val)(\vbe^T \nu_+\,\vbe) -
          (\val^T \nu_+\, \vbe)^2\,.\label{Sigsq}
\eea

  We now apply the transformation (\ref{U1}), with $\cos2\psi$ and
$\sin2\psi$ given by 
(\ref{cossinpsi}),
to the expressions (\ref{alpha0})--(\ref{Delta0}) 
in order to add in the 8th charge.  It is useful to note that for any 
symmetric $4\times 4$ matrix $X$ we shall therefore have
\bea
\val^T X \,\val  &\longrightarrow&
\ft12(\val^T X\, \val +\vbe^T X\, \vbe) +\fft{G(X)}{2\Xi}\,,\nn\\
\vbe^T X \,\vbe  &\longrightarrow&
\ft12(\val^T X\, \val +\vbe^T X\, \vbe) - \fft{G(X)}{2\Xi}\,,
\label{quad7to8}\\
\val^T X\,\vbe &\longrightarrow& 
  \fft{1}{\Xi}\, \Big[ 
   (\val^T X\,\vbe)(\val^T \nu_+\, \val -\vbe^T \nu_+\,\vbe)
  -(\val^T \nu_+\,\vbe)(\val^T X\, \val -\vbe^T X\,\vbe)\Big]\,,\nn
\eea
where
\be
G(X)= (\val^T X\,\val-\vbe^T X\, \vbe)
   (\val^T\nu_+\,\val -\vbe^T \nu_+\,\vbe) +
4 (\val^T X\,\vbe)(\val^T\nu_+\,\vbe)\,.\label{GXdef}
\ee
Note that $G(\nu_+)=\Xi^2$.

Applying these results in (\ref{alpha0})--(\ref{Delta0}), we therefore
find that the general 8-charge expressions of $\alpha$, $\beta$, $\gamma$
and $\Delta$ are given by
\begin{eqnarray}
\alpha^2& =& \val^T \nu_+ \,\val + \vbe^T \nu_+\,\vbe 
   + 2 \Sigma \,, \label{alpha1}\\
2\beta&=&  \val^T L^{-1}\, \val + \vbe^T L^{-1}\, \vbe +
  2\Sigma \,, \label{beta1}\\
2\alpha \gamma&=&(\val^T L^{-1}\,\val + \vbe^T L^{-1}\,\vbe)\,\Sigma 
 -2 (\val^T L^{-1}\,\vbe)(\val^T \nu_+\,\vbe)  \nn\\
&& + (\val^T L^{-1}\,\val)(\vbe^T \nu_+\,\vbe) +
    (\vbe^T L^{-1}\,\vbe)(\val^T \nu_+\,\val) \,,\label{gamma1}\\
4\Delta  &=&(\val^T L^{-1}\,\val)(\vbe^T L^{-1}\,\vbe)- 
  (\val^T L^{-1}\, \vbe)^2\,.\label{Delta1}
\end{eqnarray}

   We may now observe that these 8-charge expressions may be written in
a more compact notation by introducing the $SL(2,\R)$-valued 8-charge
vector $\vv^a$, where
\be
\vv^1 = \val\,,\qquad \vv^2=\vbe\,.\label{vvecdef}
\ee
We also introduce the $SL(2,\R)$-invariant antisymmetric tensor 
$\varepsilon_{ab}$, with $\varepsilon_{12}=1$, and the Kronecker delta 
$\delta_{ab}$, which is invariant only under the $U(1)$ subgroup of
$SL(2,\R)$.  Using these, the quantities $\alpha$, $\beta$, $\gamma$
and $\Delta$ in (\ref{alpha1})--(\ref{Delta1}) may be rewritten as
\bea
\alpha^2 &=& \delta_{ab}\, \vv^{a\,T} \nu_+\, \vv^b + 2\Sigma\,,\label{alpha2}\\
2\beta &=& \delta_{ab}\, \vv^{a\,T} L^{-1}\, \vv^b + 2\Sigma\,,\label{beta2}\\
2\alpha\gamma&=& \delta_{ab}\, (\vv^{a\,T} L^{-1}\, \vv^b)\, \Sigma +
  \varepsilon_{ac}\, \varepsilon_{bd}\,
  (\vv^{a\, T} L^{-1}\,\vv^b)(\vv^{c\, T}\nu_+\,\vv^d)\,,\label{gamma2}\\
8\Delta &=& \varepsilon_{ac}\,\varepsilon_{bd}\, 
   (\vv^{a\, T} L^{-1}\,\vv^b)(\vv^{c\, T} L^{-1}\,\vv^d)\,,\label{Delta2}
\eea
where
\be
\Sigma^2 =\ft12 \varepsilon_{ac}\,\varepsilon_{bd}\,
    (\vv^{a\, T} \nu_+\,\vv^b)(\vv^{c\, T} \nu_+ \,\vv^d)\,.\label{Sig2}
\ee

\bigskip

\noindent{\underline{Introducing the asymptotic scalars}:}
\medskip

  This final step is achieved by acting with the cosets $O(2,2)/U(1)^2$
and $SL(2,\R)/U(1)$, appropriately parameterised in terms of the
asymptotic values of the scalar fields.  This can be done by using
a vielbein formulation for both the $O(2,2)/U(1)^2$ 
scalar coset matrix and the $SL(2,\R)/U(1)$ scalar coset matrix. 

  For $O(2,2)/U(1)^2$, the scalar matrix $M$ given in (\ref{O22Mdef}) can
be written as
\be
M=\cV^T\,\cV\,,\qquad \cV= 
           \begin{pmatrix} E^{-1} & - E^{-1} \, \mbB \cr
                            0 & E^T\end{pmatrix}\,,\label{cVdef}
\ee
where $E$ is the zweibein for the internal 2-torus metric $\mbG$. 
Note that $\cV^T L^{-1}\, \cV=L^{-1}$. In terms of indices we have
\be
M^{ij}= \delta^{k\ell}\, \cV_k{}^i\, \cV_\ell{}^j\,,\qquad
L^{ij}= L^{k\ell}\, \cV_k{}^i\, \cV_\ell{}^j\,.
\ee
We shall then make the $O(2,2)/U(1)^2$ transformation
$\boldsymbol{\cF}\rightarrow (\overbar\cV^T)^{-1}\, \boldsymbol{\cF}$ on the
gauge fields, which implies the transformations
\be
\val \longrightarrow \overbar\cV\, \val\,,\qquad \vbe\longrightarrow
  \overbar\cV\,\vbe\,,
\ee
on the charges, where $\overbar\cV$ denotes the asymptotic value of
the scalar vielbein $\cV$.  Since $\cV^T L^{-1}\,\cV=L$ and 
$\cV^T\, \cV=M$, this means that the only change in the
expressions (\ref{alpha1})--(\ref{Sig2}) will be that the matrix 
$I_4$ in $\nu_+
= I_4+L^{-1}$ will change to $\overbar M$, and so 
\be
\nu_+\longrightarrow \mu_+\,,\qquad \mu_+= \overbar M +L^{-1}\,.\label{mupdef}
\ee

  In a similar way, the scalars $\Phi$ and $\Psi$ 
enter the $SL(2,\R)/U(1)$ scalar coset in the form
\be
{\cal N}=  {\rm e}^{2\Phi}
\begin{pmatrix} 1 &  -\Psi\cr
-\Psi& \Psi^2+{\rm e}^{-4\Phi } \end{pmatrix}\,,\label{cNdef}
\ee 
and this can be written in terms of a vielbein $U=\cU$ as
\be
\cN= \cU^T\, \cU\,,\qquad
            \cU=\begin{pmatrix} e^\Phi & -\Psi\, e^\Phi\cr 
                                0 & e^{-\Phi}\end{pmatrix} \,.
    \label{cUdef}
\ee
The components $\cU^a{}_b$ of this matrix obey the $SL(2,\R)$ invariance
condition
\be
\cU^c{}_a\, \cU^d{}_b\, \varepsilon_{cd}=\varepsilon_{ab}\,,
\ee
while, contracted instead with $\delta_{cd}$, we have
\be
\cU^c{}_a\, \cU^d{}_b\, \delta_{cd}= \cN_{ab}\,,
\ee
where $\cN_{ab}$ are the components of the $SL(2,\R)/U(1)$ scalar 
matrix (\ref{cNdef}).  Thus if we transform the electric and magnetic
charges under the $SL(2,\R)/U(1)$ transformation
\be
\vv^a\longrightarrow \overbar\cU^a{}_b\, \vv^b\,,
\ee
where $\overbar\cU^a{}_b$ is the scalar vielbein (\ref{cUdef}) with the
scalars set equal to their asymptotic values, then the only change
in the expressions 
(\ref{alpha2})--(\ref{Delta2}) and (\ref{Sig2}) will be that the
Kronecker delta $\delta_{ab}$ will be replaced by the corresponding
$SL(2,\R)/U(1)$ scalar matrix $\overbar\cN_{ab}$, where the scalars in
$\cN_{ab}$ defined in (\ref{cNdef}) are set equal to their asymptotic
values.

  Pulling together the threads of the previous discussion, if we act with
$O(2,2)/U(1)^2$ and $SL(2,\R)/U(1)$ coset elements to introduce asymptotic
values for all the scalar fields, the final expressions for the
quantities $\alpha$, $\beta$, $\gamma$ and $\Delta$ given previously in
(\ref{alpha2})--(\ref{Delta2}) will be given by
\bea
\alpha^2 &=& \overbar\cN_{ab}\, \vv^{a\,T} \mu_+\, \vv^b + 
   \Sigma\,,\label{alphag}\\
\beta &=& \ft12 \overbar\cN_{ab}\, \vv^{a\,T} L^{-1}\, \vv^b + 
    \ft12 \Sigma\,,\label{betag}\\
\gamma&=& \fft1{2\alpha}\,
\Big[\overbar\cN_{ab}\, (\vv^{a\,T} L^{-1}\, \vv^b)\, \Sigma +
  \varepsilon_{ac}\, \varepsilon_{bd}\,
  (\vv^{a\, T} L^{-1}\,\vv^b)(\vv^{c\, T}\mu_+\,\vv^d)\Big]\,,\label{gammag}\\
\Delta &=& \ft18 \varepsilon_{ac}\,\varepsilon_{bd}\,
   (\vv^{a\, T} L^{-1}\,\vv^b)(\vv^{c\, T} L^{-1}\,\vv^d)\,,\label{Deltag}
\eea
where
\be
\Sigma^2 =\ft12 \varepsilon_{ac}\,\varepsilon_{bd}\,
    (\vv^{a\, T} \mu_+\,\vv^b)(\vv^{c\, T} \mu_+ \,\vv^d)\,.\label{Sigg}
\ee
As was first observed in \cite{cvettsey}, the quartic invariant
$\Delta$ does not depend on the values of the asymptotic scalars.

Having obtained these general expressions in the heterotic formulation
for the coefficients $\alpha$,
$\beta$, $\gamma$ and $\Delta$ for static BPS black holes with
eight independent charges and arbitrary
asymptotic values for the six scalar fields, it is instructive to compare
them with the analogous expressions in the U-duality formulation, which
we obtained in (\ref{albegascal}) and (\ref{Delten}).  After making use
of the mappings (\ref{hetbetoUdual}) and (\ref{hetalUdual}) 
between the charges $\val$ and $\vbe$ of the heterotic formulation and 
the charges $(P^i,Q_i)$
of the U-duality formulation, and also the mapping (\ref{hetscalUdual})
between the scalar fields in the two formulations, it is straightforward to
verify that the two sets of expressions for $\alpha$, $\beta$, $\gamma$ and
$\Delta$ agree.

There is, however, one respect in which the two sets of expressions for
the constants $\alpha$, $\beta$ and $\gamma$ ostensibly differ in the two
formulations.  In the U-duality formulation, the $SL(2,\R)^3$-invariant 
expressions in (\ref{albegascal}) have the feature that $\alpha^2$, $\beta$
and $\alpha\gamma$ are manifestly polynomial in the eight charges. By 
contrast, in the heterotic formulation the corresponding 
$O(2,2)\times SL(2,\R)$-invariant expressions
(\ref{alphag})--(\ref{gammag}) for $\alpha^2$, $\beta$ and $\alpha\gamma$ 
are not
{\it manifestly} polynomial in the eight charges, because they are written 
using $\Sigma$, which is defined via the expression (\ref{Sigg}) for
$\Sigma^2$.  In fact one finds that $\Sigma^2$ defined in (\ref{Sigg})
turns out to be a perfect square when one evaluates it, and so
$\Sigma$ is indeed a quadratic polynomial in the charges.  However, one 
cannot write $\Sigma$ itself as a manifestly 
$O(2,2)\times SL(2,\R)$-invariant quadratic polynomial in the charges
and asymptotic scalars.
The explanation for this phenomenon turns out to be related to the
observation made earlier, in footnote \ref{Kfoot}, namely that there exists
a three-dimensional vector space of matrices that are invariant
under the $U(1)\times U(1)$ subgroup of $O(2,2)$ while being non-invariant
under the full $O(2,2)$ group.  

   As we proceeded through the steps described above, we first promoted the
5-charge expressions for $\alpha$, $\beta$, $\gamma$ and $\Delta$ to
7-charge expressions.  At that stage, we could have used the
$U(1)^2$-invariant matrix (\ref{K1K2})
\be
K= K_1+K_2 = \begin{pmatrix}0&1&0&1\cr
                            -1&0&-1&0\cr
                            0&1&0&1\cr
                            -1&0&-1&0\end{pmatrix}
\ee
in order to write $\overbar\Sigma$ in (\ref{barSigsq}) directly, unsquared,
as
\be
\overbar\Sigma= -\val^T K\,\vbe\,,\label{barSig}
\ee
as may readily be verified.  This 7-charge expression can then be augmented
to an 8-charge expression by making the replacement (\ref{U1}), with
$\psi$ given by (\ref{psi}).  Because the matrix $K$ is antisymmetric, this
replacement leaves (\ref{barSig}) unchanged, and so in the 8-charge
expressions for $\alpha$, $\beta$, $\gamma$ and $\Delta$ in
(\ref{alpha1})--(\ref{Delta1}), the expression for $\Sigma$ following
from (\ref{Sigsq}) can instead be replaced directly by the unsquared expression
\be
\Sigma=  -\val^T K\,\vbe\,.\label{Sig}
\ee
Finally, when the asymptotic scalars are introduced by acting with the
$O(2,2)/U(1)^2$ and $SL(2,\R)/U(1)$ coset matrices (\ref{cVdef})
and (\ref{cUdef}) as we did previously, the expression (\ref{Sig}) for
$\Sigma$ will become
\be
\Sigma= \ft12 \epsilon_{ab}\, \vv^{a\, T} \overbar P\,\vv^b\,,
\ee
where 
\be
\overbar P = \overbar\cV^T K\, \overbar\cV\,.
\ee
Thus we have 
\be
\Sigma= -K^{k\ell}\, \overbar\cV_k{}^i\, \overbar\cV_\ell{}^j\, 
    \alpha_i\, \beta_j 
   = - (\overbar\cV_1{}^i + \overbar\cV_3{}^i)
     (\overbar\cV_2{}^j + \overbar\cV_4{}^j)
     (\alpha_i\, \beta_j -\alpha_j\, \beta_i)\,.
\ee

\subsection{Scalar fields in the heterotic formulation}

  The derivation of the expressions for the scalar fields in the
heterotic formulation proceeds in an analogous fashion.  Here, we shall
just present the results for the $(\Phi,\Psi)$ dilaton/axion pair, associated
with the S-duality in the heterotic formalism.  They are given in the
5-charge special case (\ref{5chargehet}) by
\be
e^{-2\Phi} = \frac{V(r)^{\frac{1}{2}}}{D(r)}\,,\qquad
\Psi = \frac{\delta }{ D(r)} \,,
\label{axiodil}
\ee
where  
\be
 D(r)=r^2 +  \alpha_D\, r + \beta_D\,, 
\ee
with 
\be
\alpha_D =p^2+p^4\,,\qquad \beta_D=p^2\, p^4\,,\qquad
\delta= \ft12 q\, (p^2-p^4)\,.
\ee
As in the corresponding derivation in the U-duality formulation, we shall 
assume the asymptotic values of the scalar fields are all zero until the
final stage in the calculation.

  The augmentation to a 7-charge solution is accomplished by acting with
the $O(2)\times O(2)$ subgroup of the $O(2,2)$ T-duality, and this results in
expressions for $\alpha_D^2$, $\beta_D$ and $\delta$ that are $O(2)\times O(2)$
invariant:
\bea
{\alpha_D^2}_{\sst{(7)}} &=& (p^1+p^3)^2 + (p^2+p^4)^2
   =\vbe^T \nu_+\,\vbe \,,\nn\\
{\beta_D}_{\sst{(7)}} &=& p^1\, p^3 + p^2\, p^4 = \ft12 \vbe^T L^{-1}\, \vbe
\,,\nn\\
\delta_{\sst{(7)}} &=& -\ft14 [(p^1-p^3)(q_1-q_3)+ (p^2-p^4)(q_2-q_4)]=
 \ft12 \val^T L^{-1}\, \vbe\,.\label{abd7}
\eea
These quantities are further augmented to 8-charge expressions by 
acting with the remaining $U(1)\in SL(2,\R)$ symmetry, with angle $\psi$
given by (\ref{psi}).  Thus the right-hand-most expressions in (\ref{abd7})
are replaced according to the rules (\ref{quad7to8}), leading to
the 8-charge expressions
\bea
\alpha_D^2 &=&
\fft12\Big(\val^T \nu_+\,\val + \vbe^T \nu_+\,\vbe - \Xi\Big)\,,\nn\\
\beta_D &=&\fft14\Big(\val^T L^{-1}\,\val + \vbe^T L^{-1} \,\vbe 
               -\fft{\sigma}{\Xi}\Big)\,,\\
\delta &=& \fft1{2\Xi}\Big[ 
(\val^T\nu_+\,\vbe)(\val^T L^{-1}\, \val -\vbe^T L^{-1}\, \vbe)
-(\val^T L^{-1}\,\vbe)(\val^T \nu_+\, \val -\vbe^T \nu_+\, \vbe)\Big]\,,\nn
\eea
where $\Xi$ is given by (\ref{hetXidef}) and (\ref{Sigsq}), and
\be
\sigma= G(L^{-1})=(\val^T L^{-1}\,\val-\vbe^T L^{-1}\, \vbe)
   (\val^T\nu_+\,\val -\vbe^T \nu_+\,\vbe) +
4 (\val^T L^{-1}\,\vbe)(\val^T\nu_+\,\vbe)\,. 
\ee
These expressions can be written more compactly in terms of the
8-component charge vector $\vv^a$ defined in (\ref{vvecdef}), giving
\bea
\alpha_D^2 &=& \fft12 
    \Big(\delta_{ab}\, \vv^{a\, T}\nu_+\, \vv^b -\Xi\Big)\,,\nn\\
\beta_D &=& \fft14 \Big(\delta_{ab}\, 
               \vv^{a\, T}  L^{-1}\, \vv^b -\fft{\sigma}{\Xi}\Big)\,,\nn\\
\delta&=& -\fft1{2\Xi}\, \delta_{ac}\, \varepsilon_{bd}\,
(\vv^{a\, T} L^{-1}\, \vv^b) (\vv^{c\, T} \nu_+\, \vv^d)\,,
\label{alDbeDdelhet0}
\eea
with $\Xi$ and $\sigma$ given by
\bea
\Xi^2 &=&(\delta_{ab}\, \vv^{a\,T}\nu_+\, \vv^b)^2 - 4\Sigma^2\,,\nn\\
&=& (\delta_{ab}\, \delta_{cd}- 2\varepsilon_{ac}\, \varepsilon_{bd})\,
(\vv^{a\, T} \nu_+\,\vv^b)(\vv^{c\, T} \nu_+\, \vv^d)\,,\nn\\
\sigma&=&(\delta_{ab}\, \delta_{cd} -2\varepsilon_{ac}\, \varepsilon_{bd})
(\vv^{a\, T} \nu_+\,\vv^b)(\vv^{c\,T} L^{-1}\, \vv^d)\,.
\eea
Thus $\alpha_D^2$, $\beta_D$ and $\delta$ are all written for the general
8-charge configuration, in forms that are manifestly invariant under the
$U(1)^3$ subgroup of the full $O(2,2)\times SL(2,\R)$ duality group.

  As in the analogous earlier discussion in the U-duality formulation,
here the dilaton/axion pair $(\Phi,\Psi)$ transforms under the 
$U(1)$ subgroup of the $SL(2,\R)$ S-duality group, and so to obtain the
expressions for $\Phi$ and $\Psi$ after the augmentation from the 5-charge
seed solution to the general 8-charge solution, we should transform 
these too, according to (\ref{sdual}) with $a=d=\cos\psi$ and $b=-c=\sin\psi$,
where the $U(1)$ angle is given by (\ref{psi}).  Thus, analogously to 
(\ref{phi1chi1trans}) the dilaton and axion will then be given by
\be
 e^{-2\Phi} = \fft{\sqrt{V(r)} }{{\cal D}\, D(r)}\,,\qquad
\Psi = \fft{ {\cal C} }{ {\cal D} }\,,
\ee
where 
\bea
{\cal C}&=& \fft{\delta}{D(r)}\, \cos2\psi -
         \fft12\Big[\fft{V(r)+\delta^2}{D(r)^2} -1\Big]\, \sin2\psi\,,\nn\\
{\cal D} &=& \fft12 \Big[ 1 +\fft{V(r)+\delta^2}{D(r)^2}\Big] -
  \fft{\delta}{D(r)}\, \sin2\psi +
   \fft12\Big[1-\fft{V(r)+\delta^2}{D(r)^2}\Big]\, \cos2\psi\,.
\eea
The function $V(r)$ is given by the expression in (\ref{exm}) with
the coefficients $\alpha$, $\beta$, $\gamma$ and $\Delta$ given by
(\ref{alpha2})--(\ref{Delta2}); the function $D(r)$ is given by 
(\ref{Drdef}) with the coefficients given by (\ref{alDbeDdelhet0});
the coefficient $\delta$ is given also in (\ref{alDbeDdelhet0}); and
the expressions for $\cos2\psi$ and $\sin2\psi$ are given, as
in (\ref{cossinpsi}), by
\be
\cos2\psi= \fft1{\Xi}\, (\delta_{a1}\,\delta_{b1}-\delta_{a2}\, \delta_{b2})\,
\vv^{a\,T}\nu_+\, \vv^b\,,\qquad
\sin2\psi= 
\fft{2}{\Xi}\, \delta_{a1}\,\delta_{b2}\, \vv^{a\,T} \nu_+\,\vv^b\,.
\label{cossinc}
\ee
(Of course these last expressions should not be invariant under the $U(1)$
subgroup of the $SL(2,\R)$ S-duality, since $\psi$ is the $U(1)$ angle
of the final rotation that introduced the 8th charge.)

 Finally, the process of turning on non-vanishing asymptotic values for
the scalar fields can be accomplished by means of transformations under
the $[O(2,2)\times SL(2,\R)]/U(1)^3$ coset, as in section \ref{hetsec}.  
This means that in all the expressions above one makes the replacements
\be
\nu_+\longrightarrow \mu_+\,,\qquad
\delta_{ab} \longrightarrow \overbar\cN_{ab}\,,
\ee
where $\mu_+$ is defined in eqn (\ref{mupdef}) and $\overbar\cN_{ab}$
is obtained by setting $\Phi$ and $\Psi$ to their asymptotic values
$\overbar\Phi$ and $\overbar\Psi$ in (\ref{cNdef}).   Note that
$\delta_{a1}$ and $\delta_{a2}$ in (\ref{cossinc}) will be replaced by
$\overbar\cN_{a1}$ and $\overbar\cN_{a2}$ also.  Finally, the dilaton/axion
pair $(\Phi,\Psi)$ will transform under the $SL(2,\R)/U(1)$ coset 
also, undergoing the replacements
\be
e^{-2\Phi}\longrightarrow e^{-2\Phi-2\overbar\Phi}\,,\qquad
\Psi\longrightarrow \overbar\Psi + \Psi\, e^{-2\overbar\Phi}\,.
\ee

\subsection{Gauge fields for the 8-charge BPS black holes}

 Having obtained the expressions for the metric and the scalar fields, 
the form of the gauge fields in the general 8-charge static BPS black 
hole solutions follow straightforwardly from the gauge fields equations of
motion. 

  In the heterotic formulation we find 
\be
\cF^i = p^i\, \sin\theta\, d\theta\wedge d\varphi +
     \fft{e^{2\Phi}}{\sqrt{V}}\, M^{ij}\, 
(\alpha_j -\Psi\, \beta_j)\, dt\wedge dr\,.
\ee
As may be verified, these expressions are consistent with the 
definitions of the electric and magnetic charges in eqns 
(\ref{vecalphadef}) and (\ref{hetpdef}).

\section{Truncation to Pairwise-Equal Charges}

 The STU supergravity theory admits a consistent truncation in which 
the four gauge fields are set equal in pairs, and at the same time
two of the dilaton/axion pairs are set to zero.  Because of the triality
symmetry in the U-duality formulation, the three possible ways of 
equating pairs of gauge fields are equivalent.  We shall choose the
pairing
\be
A^1=A^2\,,\qquad A^3=A^4
\ee
in the theory described in section \ref{sl2r3sec}.  It is straightforward
to see that this truncation is consistent provided that at the same time
one sets
\be
\varphi_2=\varphi_3=0\,,\qquad \chi_2=\chi_3=0\,.
\ee
The entire bosonic Lagrangian then reduces to
\bea
{\cal L}_U &=& R\, {*\oneone} -\ft12{*d\varphi_1}\wedge d\varphi_1 - 
  \ft12 e^{2\varphi_1}\, {*d\chi_1}\wedge d\chi_1 \\
&&- 
  \fft{e^{2\varphi_1}}{1+\chi_1^2\, e^{2\varphi_1}}\,
\Big[e^{-\varphi_1}\, {*F^1}\wedge F^1 +\chi_1\, 
  F^1\wedge F^1\Big] -e^{-\varphi_1}\, {*F^3}\wedge F^3 +
 \chi_1\, F^3\wedge F^3\,.\nn
\eea
Note that this Lagrangian is invariant under the discrete symmetry
\be
A^1\longrightarrow A^3 \,,\qquad A^3\longrightarrow A^1\,,\qquad
\tau_1\longrightarrow -\fft1{\tau_1}\,,\label{Z2sym}
\ee
where as usual $\tau_1=\chi_1 + \im e^{-\varphi_1}$,
that is,
\be
e^{-\varphi_1}\longleftrightarrow \fft{e^{\varphi_1}}{
    (1+\chi_1^2\, e^{2\varphi_1})}\,,
\qquad \chi_1\longleftrightarrow -\fft{\chi_1\, e^{2\varphi_1}}{
(1+\chi_1^2\, e^{2\varphi_1})}\,.\label{Z2sym2}
\ee

In the heterotic formulation, as can be seen from (\ref{cGtoU}) and 
(\ref{cFtoU}), the corresponding truncation amounts to setting
\be
\cA^1=\cA^3\,,\qquad \cA^2=\cA^4\,,
\ee
together, again, with 
\be
\varphi_2=\varphi_3=0\,,\qquad \chi_2=\chi_3=0\,.\label{scal23zero}
\ee
The bosonic Lagrangian in the heterotic formulation, described in 
section \ref{heteroticsec}, therefore reduces to 
\bea
{\cal L}_H &=& R\, {*\oneone} - 2{*d\Phi}\wedge d\Phi - \ft12 
  e^{4\Phi}\, {*d\Psi}\wedge d\Psi \nn\\
&&-
e^{-2\Phi}\, ({*\cF^1}\wedge \cF^1 + {*\cF^2}\wedge \cF^2) + 
\Psi\, (\cF^1\wedge \cF^1 + \cF^2\wedge \cF^2)\,.
\eea

The general static BPS black hole solutions in the truncated theory are
obtained by setting the charges associated with the corresponding pairs
of equated fields to be equal also.  Thus in the U-duality formulation we
set
\be
P^1=P^2\,,\qquad Q_1=Q_2\,,\qquad P^3=P^4\,,\qquad Q_3=Q_3\,.\label{pairPQ}
\ee
In the heterotic formulation, this corresponds to setting 
\be
p^1=p^3\,,\qquad q_1=q_3\,,\qquad p^2=p^4\,,\qquad q_2=q_4\,.
\ee
Not only is the STU theory greatly simplified in the
pairwise-equal truncation of the fields, as we saw above, but also the
form of the black hole solutions becomes considerably simpler. In 
particular, the quartic metric function $V(r)$ (\ref{exm}) now
becomes a perfect square,
The metric function $V$ in eqn (\ref{exm}) also becomes a perfect square:
\be
V(r) = \Big( r^2 +\ft12 \alpha\, r +\Delta^{1/2}\Big)^2\,,
\ee
with $\alpha$ and $\Delta$ now given by
\be
\alpha=2\sqrt{(P^1+P^3)^2 + (Q_1+Q_3)^2}\,,\qquad
\Delta= (P^1\, P^3 + Q_1\, Q_3)^2\,.
\ee
(For simplicity, we first consider the case where 
the asymptotic values of the scalar fields are taken to be zero
here.)

The scalar fields $(\varphi_1,\chi_1)$ 
themselves can be read off from eqns (\ref{phi1chi1res}) and (\ref{bcdef}), 
together with the triality-related expressions for $(\varphi_2,\chi_2)$ and
$(\varphi_3,\chi_3)$ as detailed in (\ref{swap12}) and (\ref{swap13}),
after making the pairwise-equal specialisation (\ref{pairPQ}). These
latter expressions reproduce the vanishing of the $(\varphi_2,\chi_2)$ and
$(\varphi_3,\chi_3)$ as in (\ref{scal23zero}), and the former 
give (\ref{phi1chi1res})
\bea
e^{\varphi_1}&=& \fft{r^2 + b_1\, r + b_0}{r^2+ d_1\, r + d_0}\,,\qquad
\chi_1 = \fft{c_1\, r + c_0}{r^2+ b_1\, r + b_0}\,,\nn\\
b_1&=& \fft{4}{\alpha}\, \Big[ P^3\,(P^1+P^3) + Q_1\, (Q_1+Q_3)\Big]\,,\qquad
b_0=(P^3)^2 + (Q_1)^2\,,\nn\\
c_1&=& \fft{4}{\alpha}\, (P^1\, Q_1 - P^3\, Q_3)\,,\qquad
c_0= P^1\, Q_1 - P^3\, Q_3\,,\nn\\
d_1&=& \ft12\alpha\,,\qquad d_0=P^1\, P^3 + Q_1\, Q_3\,.\label{scal1bcd}
\eea
 
  It is worth noting that the asymmetry of the expressions 
for $\varphi_1$ and $\chi_1$ with respect to exchanging 
$(P^1,Q_1)$ and $(P^3,Q_3)$,
because of the asymmetry of $b_1$ and $b_0$ given in (\ref{scal1bcd}),
is in fact precisely consistent with the exchange symmetry of the 
pairwise-equal truncation of the STU theory.  This symmetry is
given in (\ref{Z2sym}).  One may verify that the scalar fields $\varphi_1$
and $\chi_1$ given in (\ref{scal1bcd}) have the property that
\be
\fft{e^{\varphi_1}}{
    (1+\chi_1^2\, e^{2\varphi_1})}= 
\fft{r^2+ d_1\, r + d_0}{r^2+ e_1\, r + e_0}\,,\qquad
\fft{\chi_1\, e^{2\varphi_1}}{
    (1+\chi_1^2\, e^{2\varphi_1})}=
-\fft{c_1\, r + c_0}{r^2+ e_1\, r + e_0}\,,
\ee
where 
\be
e_1= \fft{4}{\alpha}\, \Big[ P^1\,(P^1+P^3) + Q_3\, (Q_1+Q_3)\Big]\,,\qquad
e_0=(P^1)^2 + (Q_3)^2\,,
\ee
and so indeed the pairwise-equal solution is compatible with the
exchange symmetry given by (\ref{Z2sym2}). 

If the asymptotic values of the scalar fields are taken to be non-vanishing,
$\varphi_1\rightarrow \bar\varphi_1$ and $\chi_1\rightarrow\bar\chi_1$, 
it is straightforward to check using the results in 
section \ref{Uscalarsec} that the constants $b_1$, $b_0$, $c_1$ and $c_0$
in (\ref{scal1bcd}) are replaced by
\bea
b_1 &=& \fft{4}{\alpha}\, \Big\{e^{-\bar\varphi_1}\, [(P^3)^2 + Q_1^2] 
   + P^1\, P^3 + Q_1\, Q_3 +\bar\chi_1\, (P^1\, Q_1 -P^3\, Q_3) \nn\\
&&\qquad+
  \bar\chi_1\, e^{\bar\varphi_1}\, \Big[P^1\, Q_1-P^3\, Q_3 +
    \ft12 \bar\chi_1\, [(P^1)^2 + (P^3)^2 + Q_1^2 + Q_3^2]\Big]\Big\}\,,\nn\\
b_0&=& e^{-\bar\varphi_1}\, [(P^3)^2 + Q_1^2] + \bar\chi_1\, (P^1\, Q_1 - P^3\, Q_3) \nn\\
&&\qquad +
\bar\chi_1\, e^{\bar\varphi_1}\, \Big[ P^1\, Q_1-P^3\, Q_3 + 
  \ft12 \bar\chi_1\, [(P^1)^2 + (P^3)^2 + Q_1^2 + Q_3^2]\Big]\,, \nn\\
c_1 &=& \fft4{\alpha}\, \Big[ P^1\, Q_1 - P^3\, Q_3 + \bar\chi_1\, 
        e^{\bar\varphi_1} \, [(P^1)^2 + Q_3^2]\Big]\,,\nn\\
c_0&=& P^1\, Q^1 - P^3\, Q_3 + \bar\chi_1\, e^{\bar\varphi_1}\,
            [(P^1)^2 + Q_3^2]\,.
\eea
Note that $b_1=\fft{4}{\alpha}\, (b_0 + P^1\, P^3 + Q_1\, Q_3)$, and
$c_1=\fft{4}{\alpha}\, c_0$.  The quantity $\alpha$ is now given by
\bea
\alpha^2 &=& 4 e^{\bar\varphi_1}\, \Big[ (P^1 + \bar\chi_1\, Q_1)^2 +
            (Q_3 - \bar\chi_1\, P^3)^2\Big] + 8(P^1\, P^3 + Q_1\, Q_3)\nn\\
&& +
   4e^{-\bar\varphi_1}\, [(P^3)^2 + Q_1^2]\,.
\eea
Finally, the scalar fields $\varphi_1$ and 
$\chi_1$ themselves should be transformed, as in eqn 
(\ref{phichishift}). 

\section{Symmetry Transformations and Asymptotic Scalars}\label{Ksec}

In this section, we shall address a conformal transformation for 
BPS black holes, studied in \cite{borsduff2}, in the context of STU theory. 
In order to study the conformal transformation, we need to go outside 
the approach of the rest of this paper, and adopt the pre-potential 
formalism studied in \cite{sabra,lust}. In \cite{lust}, the authors discuss 
various stationary solutions of  $D=4,\,{\cal N}=2$ supergravity, among 
which the one with pre-potential 
\be
\label{ppt}
F(X^I) = -\fft{X^1X^2X^3}{X^0}
\ee
is of interest to us.  The resulting theory, as we shall see later, 
corresponds to STU supergravity in the 3+1 formulation that we discussed
in section \ref{3+1sec}.

   Let us consider a K\"ahler manifold with K\"ahler potential given by
\be
K = -\log[\,\im({\bar X^I}W_I - X^I{\bar W_I})]
\ee
where $X^I$ and $W_I$\footnote{$W_I$ is denoted as $F_I$ in most of the literature dealing with supergravity and special geometry, however we choose not to use the notation $F_I$ to avoid confusion with field strength.} are related to the ``holomorphic section'' of the underlying K\"ahler manifold via the usual definition
\be
\begin{pmatrix}
X^I\\
W_I
\end{pmatrix}
=
e^{-\fft K2}
\begin{pmatrix}
L^I\\
M_I
\end{pmatrix}\label{XWLM}
\ee
where $\begin{pmatrix}
L^I\\
M_I
\end{pmatrix}$ constitute the ``holomorphic section''. The four quantities, $W_I$, are evaluated from the pre-potential mentioned in (\ref{ppt}) using the definition: $W_I = \fft{\del F}{\del X^I}$. Physical scalars $z^A,\,A=1,2,3$, which parameterise the K\"ahler manifold, are given in terms of $X^I$ 
functions as follows
\be
z^A = \fft{X^A}{X^0},\quad A=1,2,3.
\label{zx}
\ee

It can be shown that the metric function of the 8-charge static BPS black 
hole should be related to the K\"ahler potential (since the metric has to 
be duality invariant, it should be related to a similar duality 
invariant quantity in special geometry), and $X^I$, $W_I$ and their 
complex conjugates should be given by harmonic functions. These are made 
to ensure that the bosonic solutions are supersymmetric (i.e. solutions 
which render the supersymmetry variations $\delta \psi_\mu$ and 
$\delta \lambda^A$ equal to zero). We refer to \cite{lust} for 
detailed results. 
\begin{align}
& e^{-2 U} = e^{-K} =\, \im (\bar X^IW_I - X^I\bar W_I)\,,\nn\\
& \im (X^I - {\bar X^I}) = \Hp^I, \quad \text{where} \quad \Hp^I = \hp^I + \fft{p^I}{r},\nn\\
& \im (W_I - {\bar W_I}) = \Hq_I, \quad \text{where} \quad \Hq_I = \hq_I + \fft{q_I}{r}.
\label{xxww}
\end{align}
Here $\hp^I$ and $\hq_I$ are integration constants, while $q_I$ and $p^I$ 
are the electric and magnetic charges.
Using (\ref{zx}) and (\ref{xxww}), the metric and scalar fields are 
expressed in terms of the harmonic functions below,
\bea
z^A &=& \fft{\left(2\Hp^A\Hq_A - \Hp^I\Hq_I\right) - \im e^{-2U}}{\left(d_{ABC}\Hp^B\Hp^C + 2 \Hp^0\Hq_A\right)}\,, \qquad A=1, 2, 3\,, 
   \nn\\
e^{-4U} &=& {\fft 1{r^4}}V(r) \nn\\
&=& -(\Hp^I\Hq_I)^2 + (d_{ABC}\Hp^B\Hp^C d^{ADE}\Hq_D \Hq_E) \nn\\
&& +\, 4\,\Hp^0\Hq_1\Hq_2\Hq_3 - 4\,\Hq_0\,\Hp^1\Hp^2\Hp^3\,.
\label{zmet}
\eea
$d_{ABC}$ is the ``symmetrised $\epsilon$ tensor,'' $d_{ABC}=|\epsilon_{ABC}|$,
so 
\be
d_{123} = 1 = d_{213} = \text{(any permutation of 1,2,3 indices)}\,.
\ee
Towards the end of this section we shall relate the charges, scalar fields 
and the metric mentioned in the above equation to those in 
the U-duality formulation, via the 3+1 formulation discussed in section 
\ref{3+1sec}. 

Not all the eight constants $\hq_I$ and $\hp^I$ are independent; they
are constrained by two conditions. Since the metric is asymptotically flat, 
we have the constraint
\bea
1 &=&  -(\hp^I\hq_I)^2 + (d_{ABC}\,\hp^B\hp^C d^{ADE}\,\hq_D \hq_E) \nn\\
&& +\, 4\,\hp^0\hq_1\hq_2\hq_3 - 4\,\hq_0\,\hp^1\hp^2\hp^3
\label{ct1}
\eea
The other constraint is \cite{sabra}
\be
\hp^I\, q_I - \hq_I\, p^I = 0\,,
\label{ct2}
\ee
which arises from a more fundamental requirement of special geometry: 
\bea
\langle V, {\cal D}_A V\rangle = 0, \quad \text{where} 
\quad V = 
\begin{pmatrix}
L^I\\
M_I
\end{pmatrix},
\qquad {\cal D}_A \equiv \del_A + \ft12 (\del_A K)\,,\label{DVD}
\eea
where $\del_A=\del/\del z^A$.
The angular bracket denotes the inner product with respect to the 
symplectic metric. We again refer to \cite{lust,sabra} for details. Thus only 
six out of the eight integration constants $\hp^I$ and 
$\hq_I$ are independent.

We now have all the necessary ingredients for discussing the 
conformal inversion of the static 8-charge BPS metric. Since the metric is 
in general is parameterised by eight charges and the asymptotic values of 
the six scalar fields, the conformal inversion transforms the charges 
and the asymptotic values of the scalars into a new set of charges and 
asymptotic scalar values in the conformally rescaled metric (see, for
example, section 4 in \cite{cvposa1}). Although it is relatively easy 
to find the relations between constants $\alpha$, $\beta$, $\gamma$ and $\Delta$ appearing in the metric function $V(r)$ in section 
\ref{Udualitysec} and those in the conformally rescaled metric, 
it is quite difficult to find the relation between original and the 
transformed charges and scalars. It is nonetheless possible to circumvent 
this difficulty by viewing the metric as being parameterised by $q_I,\,p^I,\,\hq_I$, and $\hp^I$. It should again be emphasised that out of these 
sixteen parameters, only fourteen (i.e. eight charges and six out of the eight $\hq_I$ and $\hp^I$ constants) are independent, thus giving us the same 
number of parameters as we have when the metric is written in terms  
of the eight charges and the asymptotic values of the six
scalars. Following \cite{borsduff2}, if we implement the transformation
\bea
\begin{pmatrix}
\hq_I \\ \hp^I
\end{pmatrix}
\rightarrow
\begin{pmatrix}
{\hat \hq}_I\\{\hat \hp}^I
\end{pmatrix}
= \Delta^{-\fft 14} \begin{pmatrix}
q_I \\ p^I
\end{pmatrix},
\qquad
\begin{pmatrix}
q_I \\ p^I
\end{pmatrix}
\rightarrow
\begin{pmatrix}
{\hat q}_I\\{\hat p}^I
\end{pmatrix}
= \Delta^{\fft 14} \begin{pmatrix}
\hq_I \\ \hp^I
\end{pmatrix},
\label{cinv}
\eea
together with $r \rightarrow {\hat r} = \sqrt{\Delta}/r$, it is 
easy to check that we obtain the correct conformally rescaled metric,
namely
\bea
&&V(r) = \fft{\Delta}{{\hat r}^4} {\hat V}({\hat r}) ,
\qquad ds^2 = \fft{\sqrt{\Delta}}{{\hat r}^2}{\hat ds}^2\nn\\
&&{\hat ds}^2 = -\frac{\hat r^2}{\sqrt{{\hat V}(\hat r)}}\, dt^2 + \frac{\sqrt{{\hat V}(\hat r)}}{\hat r^2}\, (d\hat{r}^2 + \hat{r}^2 
d\Omega^2)\,,
\eea
where $\hat V({\hat r})$ is obtained from the second equation 
in (\ref{zmet}) by replacing $\Hp^I$ and $\Hq_I$ by 
${\hat \Hp}^I$ and ${\hat \Hq}_I$, with ${\hat \Hp}^I = {\hat \hp}^I + 
{\hat p}^I/{\hat r}$, etc.\footnote{In \cite{borsduff2} and in \cite{cvposa1}
the quantities after the conformal inversion were denoted using tildes. Here,
we are instead using hats to denote the conformally-inverted quantities,
reserving tildes for the functions $\wtd H^I=\td h^I +p^I/r$ in the notation
of \cite{lust}.}

It is now instructive to see what happens to the asymptotic values of the scalar fields under these transformations. Under the transformations in (\ref{cinv}), 
\bea
\begin{pmatrix}
{\hat \Hp}^I \\ {\hat \Hq}_I
\end{pmatrix} = r\, \Delta^{-\fft 14}
\begin{pmatrix}
\Hp^I \\ \Hq_I
\end{pmatrix},\qquad 
e^{-2{\hat U}} = r^2 \Delta^{-\fft 12} e^{-2U}\,.
\eea
According to (\ref{zmet}), this implies for scalars that
\be
{\hat z}^A \left({\hat r},{\hat q}_I,{\hat p}^I,
{\hat \hp}^I,{\hat \hq}_I \right) = {z}^A 
\left(r, q_I, p^I, \hp^I, \hq_I \right).
\ee
Thus the functional form of the scalar fields remain unchanged under the conformal transformation. This implies  that 
\bea
\lim_{r\rightarrow\infty} z^A = 
\frac{(2\hp^A\hq_A -\, \hp^I\hq_I) - \im}{d_{ABC}\,\hp^B\hp^C + 
2\,\hp^0\hq_A} = \lim_{{\hat r}\rightarrow 0}{{\hat z}^A} 
  = \frac{2\, {\hat p}^A {\hat q}_A - {\hat p}^I {\hat q}_I 
- \im\, \sqrt{\hat V(\hat p^I,\hat q_I)}}{d_{ABC}\, {\hat p}^B 
           {\hat p}^C + 2\, {\hat p}^0 {\hat q}_A}\,,
\eea
where ${\hat V}(\hat p^I,\hat q_I)$ is the metric function at 
the horizon after inversion, which can be obtained by the replacements
$\Hp^I \rightarrow {\hat p}^I$ and $\Hq_I\rightarrow {\hat q}_I$ in the 
lower equation in (\ref{zmet}). In other words, the above equation implies that 
the asymptotic values of the scalar fields in the original metric map to 
the values of the fields at the horizon of the transformed metric. 

 It is important to distinguish this property of the scalar fields 
undergoing conformal inversion from that considered in \cite{cvposa1}. 
In \cite{cvposa1}, the asymptotic values of the scalar fields were taken to
be be zero in the static black hole solution of STU supergravity, and it 
was mandated that after the conformal inversion the metric should again 
be an 8-charge solution with vanishing asymptotic values for the 
scalar fields.  By contrast, in the present discussion, if we start with zero asymptotic values for the scalar fields, we will end up with the scalar fields becoming \textit{zero at the horizon of the inverted metric}, but at spatial infinity the \textit{scalar fields will be non-zero}. Therefore, the type of transformation considered in (\ref{cinv}) clearly does not satisfy the 
requirements of the conformal inversion discussed in \cite{cvposa1}, 
which involved mapping any member of the restricted 
8-parameter class of charged
black holes with vanishing asymptotic scalars to another member of this
8-parameter class.  Instead it
provides an alternative way of implementing the inversion,
describing a mapping from any member of the 14-parameter general class of
black hole solutions to another member of the class.

Although we have in principle demonstrated the effect that the conformal inversion has on the metric, the charges, the constants $\hp^I$, $\hq_I$, 
and the scalars, the discussion might appear abstract to some degree since no 
real connection has been made between a particular theory and the particular 
pre-potential considered in (\ref{ppt}). To establish that connection, we need 
to look at two equations which arise as a consequence of holomorphicity of the 
section $(L^I,M_I)$ of the underlying K\"ahler manifold, namely,
\bea
W_I = {\cal N}_{IJ}X^J,\qquad {\cal D}_A W_I = 
\overbar{\cal N}_{IJ}\,{\cal D}_A X^J\,,
\label{nij}
\eea
where ${\cal N}_{IJ}$ is the complex matrix describing the coupling
of the scalars to the field strengths (see below).
Here ${\cal D}_A = \del_A + (\del_A K)$ is the appropriate 
covariant derivative with the 
K\"ahler connection.  Note that the change of the coefficient of the 
$(\del_A K)$ term from $\ft12$ in eqn (\ref{DVD}) to $1$ here is because the
different weights of the differentiands with respect to $K$, as can be seen
in eqn (\ref{XWLM}).   

   Since we know the relation between $X^A$ and the scalars $z^A$ 
via (\ref{zx}), it is possible to solve the $(4+12) = 16$ conditions 
coming from (\ref{nij}) to determine ${\cal N}_{IJ}$ in terms of $z^A$. 
The gauge field part of the Lagrangian in the 3+1 formulation 
considered in section \ref{3+1sec} has this structure, and if we choose 
\be
z^A = -\chi_A - i e^{-\varphi_A}
\ee
then it is straightforward to see that the Lagrangian in (\ref{lag31}) 
becomes
\be
{\cal L} = R\, + {\cal L}(\varphi,\chi) -\fft12 \Im(N_{IJ})*F^I\wedge F^J + 
\fft12 \Re (N_{IJ})F^I\wedge F^J\,,\label{LF}
\ee
where scalar kinetic terms are given by
\bea
{\cal L}(\varphi,\chi) = -2 g_{A{\bar B}}*dz^A\wedge d{\bar z}^{\bar B} 
= -\fft12 \left(\sum_{i=1}^3 *d\varphi_i\wedge d\varphi_i + 
e^{2\varphi_i}*d\chi_i\wedge d\chi_i \right)\,,
\eea
with $g_{A{\bar B}} =\del_A\del_{\bar B}\, K$.

Finally, the metric considered in (\ref{zmet}) as a general function of 
$\Hp^I$ and $\Hq_I$ can be shown to be equal to the static 8-charge metric 
in the U-duality formulation upon relating the charges in this section to 
the charges of U-duality frame via the mapping\footnote{The field strengths
$F_{\sst{(BLS)}}$ in \cite{lust} are normalised differently from ours,
with $F_{\sst{(BLS)}}=1/(2\sqrt2)\, F$.  The Lagrangian (\ref{LF}) is
written in terms of our convention for the field strength normalisation.  Comparison of the expressions for
the fields in the black hole solutions in \cite{lust} with our expressions leads to 
the relations in eqn (\ref{qQmap}).}
\bea
&&p^A = \frac{P^A}{\sqrt{2}}, \quad q_A = 
\frac{Q_A}{\sqrt{2}}, \quad A=1,2,3.\nn\\
&&p^0 = \frac{Q^4}{\sqrt{2}},\quad q_0 = -\frac{P^4}{\sqrt{2}}\,.
\label{qQmap}
\eea
Note that it would also be natural, when mapping from the (3+1) formulation
of STU supergravity used in \cite{lust} to the U-duality formulation, to
relabel the constants $\tilde h^0$ and $h_0$ in the same way as $p^0$ and
$q_0$ are relabelled in (\ref{qQmap}):
\be
\tilde h^0 = h_4\,,\qquad h_0=-\tilde h^4\,.\label{0to4redef}
\ee
Thus the functions $\wtd H^I$ and $H_I$ with $I=0,1,2,3$ would also now
be defined instead for $I=1,2,3,4$, with
\be
\wtd H^I = \td h^I + \fft{P^I}{\sqrt2 \,r}\,,\qquad
H_I= h_I + \fft{Q_I}{\sqrt2\, r}\,,\qquad I=1,2,3,4\,.
\ee

Although we have given a way to write the metric in terms of the 
constants $\hp^I$, $\hq_I$ and the charges, and shown how these should 
transform under conformal inversion, it should be emphasised that 
the quantities $\hp^I$ and $\hq_I$ are not all physical and independent. 
In order to evaluate 
them in terms of physical charges and asymptotic scalar values, we need 
to use the first equation in (\ref{zmet}). Keeping in mind the equation (\ref{zx}), we could write six equations for $\hp^I$ and $\hq_I$:
\bea
0 &=& 2\hp^A\hq_A - \hp^I\hq_I + \left(d_{ABC}\hp^B\hp^C 
  + 2\hp^0\hq_A\right){\bar\chi}_A, \qquad A=1,2,3.\nn\\
f_A &=& d_{ABC}\hp^B\hp^C + 2 \hp^0\hq_A, \qquad A=1,2,3\,,
\eea
where ${\bar\chi}_A$ and $f_A\equiv e^{\bar\varphi_A}$ are the asymptotic values of the scalars. These six equations, along with (\ref{ct1}) and (\ref{ct2}), enable us to solve for the eight constants, giving
\bea
\hp^0 &=& a\,,\quad \hp^1 = -a\,{\bar\chi}_1 + 
     \fft b{f_1}\,,\quad \hp^2 = -a\,{\bar\chi}_2 + 
       \fft b{f_2}\,,\quad \hp^3 = -a\,\bchi_3 + \fft b{f_3}\nn\\
\hq_0 &=& -a\,\left(\bchi_1\bchi_2\bchi_3 - 
\fft{\bchi_1}{f_2 f_3} - \fft{\bchi_2}{f_1 f_3} - 
\fft{\bchi_3}{f_1 f_2}\right) -
   b\,\left( \fft 1{f_1 f_2 f_3} + \fft{\bchi_2\bchi_3}{f_1} 
   + \fft{\bchi_1\bchi_3}{f_2} + \fft{\bchi_1\bchi_2}{f_3}\right)\,,\nn\\
\hq_1 &=& a\,\left( \fft 1{f_2f_3} + \bchi_2\bchi_3\right) + 
b\,\left(\fft{\bchi_2}{f_2} + \fft{\bchi_3}{f_3}\right)\,,\nn\\
\hq_2 &=& a\,\left( \fft 1{f_1f_3} + \bchi_1\bchi_3\right) + 
b\,\left(\fft{\bchi_1}{f_1} + \fft{\bchi_3}{f_3}\right)\,,\nn\\
\hq_3 &=& a\,\left( \fft 1{f_2f_1} + \bchi_2\bchi_1\right) + 
b\,\left(\fft{\bchi_2}{f_2} + \fft{\bchi_1}{f_1}\right)\,,\label{hsols}
\eea
where $a$ and $b$ are given by
\be
a = \fft{\sqrt{f_1f_2f_3}}{\sqrt 2} \fft{D_q}{\sqrt{D_q^2 + D_p^2}}\,,
\qquad b = \fft{\sqrt{f_1f_2f_3}}{\sqrt 2} 
   \fft{D_p}{\sqrt{D_q^2 + D_p^2}}\label{hsols2}
\ee
and 
\bea
D_q &=& q_4 + f_1f_2\left(q_1 - \bchi_2\,p^3 -\bchi_3\,p^2 - 
                \bchi_2\,\bchi_3\,q_4\right) + f_1f_3\left(q_2 - 
      \bchi_1\,p^3 - \bchi_3\, p^1 - \bchi_1\,\bchi_3 \,q_4\right)\nn\\
&& + f_1f_2\left(q_3 - \bchi_1\,p^2 - \bchi_2\,p^1 - 
\bchi_1\,\bchi_2\,q_4\right)\,,\nn\\
D_p &=& f_1f_2f_3\left(p^4 + \bchi_1\,q_1 + \bchi_2\,q_2 + 
\bchi_3\,q_3 - \bchi_2\,\bchi_3\,p^1 - \bchi_1\,\bchi_3\,p^2 - 
\bchi_1\,\bchi_2\,p^3 - \bchi_1\,\bchi_2\,\bchi_3\,q_4\right) \nn\\
&& + f_1\left(p^1 + \bchi_1\,q_4\right) + f_2\left(p^2 + 
\bchi_2\,q_4\right) + f_3\left(p^3 + \bchi_3\,q_4\right)\,.\label{hsols3}
\eea
Note that in the special case where the asymptotic values of the scalar
fields $\varphi_i$ and $\chi_i$ are taken to vanish, the $\tilde h^I$ and
$h_I$ constants become simply
\bea
&&\td h^1=\td h^2=\td h^3=\td h^4= \fft{\sum_I P^I}{\sqrt 2\,
  \sqrt{(\sum_J P^J)^2 + (\sum_J(Q_J)^2}}\,,\nn\\
&&h_1=h_2=h_3=h_4= \fft{\sum_I Q_I}{\sqrt 2\,
  \sqrt{(\sum_J P^J)^2 + (\sum_J(Q_J)^2}}\,,
\eea
(after the change to the U-duality notation in (\ref{0to4redef})).

Plugging the solutions (\ref{hsols}) 
into the metric in (\ref{zmet}), along with 
the mapping (\ref{qQmap}), one can show that the metric is indeed equal 
to the static metric of the U-duality formulation (it is most easily done 
when the asymptotic scalars are set to zero). These voluminous but 
symmetric equations enable one to find the eight constants $\hp^I$ and 
$\hq_I$ 
in terms of the eight charges and the asymptotic values of the six scalars. 
One can think of this as an ``initial value'' assignment for the eight 
$\hp^I$ and $\hq_I$ constants, prior to the inversion. One could then formulate 
the inversion problem entirely in terms of these constants and charges, forgetting about the scalars. It is always possible, after inversion, to re-express: I) the transformed charges in terms of original charges and 
scalars using (\ref{cinv}) and the aforementioned equations, 
and II) the transformed 
scalars in terms of the original charges and asymptotic values of the 
original scalars using (\ref{zmet}), after expressing (\ref{zmet}) in 
terms of ``hatted'' quantities. The latter, as we have already shown, 
turn out to be equal to the original set of scalars.

\section{Concluding Remarks and Outlook}

Studies of BPS black holes in four-dimensional ungauged supergravity 
theories, such as the extremal STU black holes of the ${\cal N}=2$ 
supergravity theory coupled to three vector supermultiplets,  
have been a subject of intense 
research  ever since their discovery \cite{cvyoII,cvettsey}. 
The study of their properties
attracted extensive efforts over years, with recent ones focusing on 
their enhanced symmetries, such as  Aretakis  and Newman-Penrose 
charges \cite{godgodpop,cvposasa}, and Couch-Torrence-type 
symmetries \cite{cvposa1}.
It is important to note that these features of extremal STU black holes 
stem from and are closely related to those of extremal 
Reissner-Nordstr\"om  and Kerr-Newman black holes of Maxwell-Einstein gravity, 
these are a special case of STU black holes. The STU BPS black holes 
are fourteen-parameter solutions, specified by four electric and four 
magnetic charges,  {\it and} by the asymptotic values of the six scalar fields. 

In the past, most analyses focused on the BPS black holes where the 
asymptotic values of scalar fields were taken to be their canonical zero 
values.  The explicit form of such black holes was constructed in  \cite{cvettsey}, in the heterotic formulation, and the analogous process
was employed recently in \cite{cvposa1} to construct the eight-charge 
solutions
in the U-duality formulation of STU supergravity.  In \cite{marrani}, 
all attractor flows for BPS and non-BPS black holes were described in 
full generality in the STU symplectic frame.

One of the main purposes of the present paper was 
to derive systematically the full 
explicit solutions, both in the U-duality and the heterotic frames, 
in sections 3 and 4 respectively \footnote{In 
\cite{cvettsey} explicit results for the mass and the horizon area of these black holes in the heterotic frame were obtained. This analysis showed 
that the horizon area, and thus the Bekenstein entropy, is 
{\it independent} of the asymptotic 
values of the scalar fields.}. The solutions are specified by
coefficients in the functions giving the metric and scalar fields that
are expressed in terms of manifestly covariant U-duality or 
heterotic-duality quantities.  In section 5 we also presented
BPS black hole solutions of the consistently truncated STU supergravity 
obtained by equating the four electromagnetic fields in pairs and
at the same time setting four of the six scalar fields to zero.
This supergravity theory, with two gauge fields and one complex 
(axion-dilaton) scalar field, and its black hole solutions, 
are significantly simplified, and are well suited to further explicit 
studies.   In section 6 we made use of some of the general 
results we obtained in order to discuss in detail a conformal
inversion symmetry of the BPS black hole solutions.

The new results we have obtained immediately lend  themselves to study 
further symmetry structures of these generalised solutions, 
such as generalisations of the conformal inversion symmetries of 
the Couch-Torrence type.  For that purpose, in section 6, we consider 
the  eight-charge static BPS black holes  \cite{lust},
formulated in the description of the theory in terms of the K\"ahler 
geometry of the scalar manifold \cite{FVP}. In this formulation the BPS 
black hole solutions are expressed in terms of eight harmonic functions, subject to two constraints \cite{sabra} and thus, again, the solution  is specified in terms of fourteen parameters. We employed a  transformation, 
discussed  in \cite{borsduff2}, by acting with an inversion of the radial
coordinate, which, together with a conformal rescaling of the metric, 
maps the horizon to infinity and 
vice versa.  As we also obtained the explicit map of the black 
hole solutions in this formulation to those in the U-duality frame, we 
can show how any member of the 
 general fourteen-parameter family of static BPS black holes
(characterised by the eight charges and  six  asymptotic scalar values) is 
mapped by this conformal inversion to another
member of the family.  This type of conformal inversion is different from, 
although closely related to, the one considered in 
\cite{godgodpop,cvposa1}, where the asymptotic values of scalar fields 
remained unchanged (actually, always set to zero) under the inversion.

Building on technical advances presented in this paper, we foresee a 
number of important future research directions. 
First of all, black hole solutions of the STU supergravity have in 
principle an immediate generalisation to black holes of maximally 
supersymmetric ungauged supergravity theories, i.e. the ${\cal N}=8$  
and the ${\cal N}=4$ supergravities of the toroidally compactified effective 
Type II and heterotic superstring theories, respectively. 
In particular, we expect that the BPS solution of ${\cal N}=4$ 
supergravity with $O(6,22)$ T-duality symmetry and $SL(2,\R )$ 
S-duality symmetry (at the level of the equations of motion) can be 
obtained in a straightforward way by acting on the STU black hole 
with a subset of the
global symmetry generators in the heterotic frame. The final solution 
will then be parameterised by 28 electric and 28 magnetic charges, 
and the asymptotic values of the 134 scalar fields in the coset of 
$O(6,22)/[O(6)\times O(22)] \times SL(2,\R )/U(1)$. It is expected 
that the final result should represent a straightforward generalisation 
of the resulting section 4, with  charge vectors  ${\vec \alpha}$ 
and $\vec\beta$ characterising the 28 electric and  28 magnetic charges,
respectively, the matrix $L$ now denoting the $O(6,22)$ invariant matrix, 
and  the matrix $\mu_+$ parameterising the corresponding  asymptotic values 
of 132 scalar fields in the $O(6,22)/[O(6)\times O(22)]$ coset. 
(Note again, that in this case  the mass and the horizon area were 
obtained in \cite{cvettsey}.)
On the other hand, ${\cal N}=8$ supergravity has an 
$E_{7,7}$ U-duality symmetry, and  the general BPS black hole solution can in principle be generated by acting on the STU black hole solution with 
a subset of $E_{7,7}$ generators. It would be of great interest to 
obtain the explicit form of the full BPS black hole solution in terms of 
manifestly $E_{7,7}$-covariant coefficients. 

   Another important application is to focus on black hole properties 
as a function of the asymptotic values of the scalar fields, which in 
effective string theory specify the moduli of the corresponding 
string compactification. The explicit expressions obtained in this paper 
would allow us to employ these black holes for studies of various 
so-called swampland conjectures \cite{vafa,ogva,arkani}. In particular, 
we plan to explore the implications of these black holes for the 
swampland distance conjecture \cite{ogva}, and its connection to the 
weak gravity conjecture \cite{arkani}.  
(The swampland distance conjecture  argues that as one moves 
toward the boundary 
of moduli space, there appears an infinite tower of states with masses
 approaching zero exponentially as a function of the traversed distance. 
The weak gravity conjecture argues that in any consistent quantum gravity 
there must exist a particle whose charge-to-mass ratio equals or exceeds 
the extremality  bound for black hole solutions of that theory.)  For 
recent work, tying together the two conjectures via special examples of 
BPS black holes, see \cite{val} and references therein. 
We would also like to emphasise that since the BPS black hole 
solutions presented in this paper account for both electric and 
magnetic charges, they would allow for probing non-perturbative effects 
in the moduli space of string compactifications, such as the appearance of
light/massless dyonic BPS states in the middle of moduli space,  where 
they could signify, for example, the appearance of enhanced gauge symmetry 
and/or supersymmetry (c.f., \cite{cvyoIII}).

 \vskip 1cm
\section*{Acknowledgments} 
The work of M.C. is supported in part by the DOE (HEP) Award DE-SC0013528, 
the Fay R. and Eugene L. Langberg Endowed Chair (M.C.) and the Slovenian 
Research Agency (ARRS No. P1-0306).  
The work of C.N.P. is supported in part by DOE 
grant DE-FG02-13ER42020.

\appendix

\section{Kaluza-Klein $T^2$ Reduction From Six Dimensions}\label{KKredsec}

  Using the notation and conventions of \cite{lupomax,cjlp1}, the
$T^2$ reduction of the six-dimensional Lagrangian (\ref{lag6}) is
accomplished by means of the reduction ans\"atze 
\bea
d\hat s_6^2 &=& e^{\ft12(\varphi_1+\varphi_3)}\, ds_4^2 +
e^{-\ft12(\varphi_1+\varphi_3)}\, \Big[ e^{-\tvp_2}\, (h^1)^2
   + e^{\tvp_2}\, (h^2)^2\Big]\,,\label{metred}\\
\hat B_\2 &=& B_\2 + A_{\1 1}\wedge dz^1 + A_{\1 2}\wedge dz^2 -
    A_{\0 12} \, dz^1\wedge dz^2\,,\label{B2red}\\
\hat\phi &=&\fft1{\sqrt2}\, (\varphi_1-\varphi_3)\,,\label{phired}
\eea
where
\be 
h^1 = dz^1 + \cA_\1^1 +\cA^1_{\0 2}\, dz^2\,,\qquad
h^2 = dz^2 + \cA_\1^2\,,
\ee
and $(z^1,z^2)$ are the coordinates on the internal 2-torus.
The six-dimensional 3-form field strength is then given by
\be
\hat H_\3=d \hat B_\2 = H_\3 + F_{\2 1}\wedge h^1 +
  F_{\2 2}\wedge h^2 - F_{\1 12}\, h^1\wedge h^2\,,\label{4fields}
\ee
with the four-dimensional field strengths $H_\3$, $F_{\2 1}$, $F_{\2 2}$
and $F_{\1 12}$ being read off by substituting (\ref{B2red}) into
(\ref{4fields}).  It turns out to be advantageous to make redefinitions
of certain of the potentials in order to obtain the four-dimensional
theory in a parameterisation in which the axionic scalars $\cA^1_{\0 2}$
from the metric and $A_{\012}$ from the $\hat B_\2$ field occur 
without derivatives in their couplings to the vector fields.  Thus we define
primed potentials as follows
\bea
A_{\1 1} = A_{\1 1}' + \cA^1_{\0 2} \,,\qquad
A_{\1 2}= A_{\1 2}' + A_{\0 12}\, {\cA_\1^1}\,,\qquad
\cA_\1^1 = {\cA_\1^1}' + \cA^1_{\0 2}\, \cA_\1^2\,.
\eea
We also define a primed potential $B_\2'$ for the four-dimensional
3-form $H_\3$ via
\be
B_\2 = B_\2' + A_{\0 12}\, {\cA_1^1}' \wedge \cA_\1^2\,.
\ee
The four-dimensional dressed field strengths are now given by
\bea
\cF_\2^1 &=& d{\cA_\1^1}' + \cA^1_{\0 2}\, d\cA_\1^2\,,\qquad
\cF_\2^2 = d\cA_\1^2\,,\nn\\
F_{\2 1} &=&d A_{\1 1}' - A_{\0 12}\, d\cA_\1^2\,,\nn\\
F_{\2 2} &=& d A_{\1 2}' + A_{\0 12}\, d{\cA_\1^1}' -
  \cA^1_{\0 2}\, d A_{\1 1}' + \cA^1_{\0 2}\, A_{\0 12}\, d\cA_\1^2\,,\nn\\
H_\3 &=& dB_\2' - dA_{\1 1}'\wedge {\cA_\1^1}' -
         dA_{\1 2}'\wedge \cA_\1^2\,.
\eea
In this form, after then dualising the 2-form potential $B_\2$ to an axion,
the bosonic STU supergravity Lagrangian was presented in \cite{chcvlupo}.

  In this paper we shall denote the four gauge potentials by $(\tA_1,\tA_2,
A^3,A^4)$, with
\be
\tA_1={\cA_\1^1}'\,,\qquad \tA_2=A_{\1 1}'\,,\qquad A^3=\cA_\1^2\,,\qquad
A^4=A_{\1 2}'\,.
\ee
We also define the axions $\tchi_2$ and $\chi_3$ by
\be
\tchi_2=-\cA^1_{\0 2}\,,\qquad \chi_3= A_{\0 12}\,.
\ee
(The tildes are placed on the potentials $\tA_1$ and $\tA_2$ to signify
that these are the fields that we shall dualise when passing to the
U-duality formulation of the STU supergravity.  The tildes on $\tchi_2$,
and on $\tvp_2$ earlier, are used because we are reserving the
symbols $\chi_2$ and $\varphi_2$ for redefined fields we shall be using later.)
The four-dimensional Lagrangian is then given by 
\bea
{\cal L} &=& R\,{*\oneone} -\ft12{*d\varphi_1}\wedge d\varphi_1 
 -\ft12 e^{2\varphi_1}\, {*d\chi_1}\wedge d\chi_1 
 -\ft12 {*d\tvp_2}\wedge d\tvp_2 -\ft12 e^{2\tvp_2}\,{*d\tchi_2}\wedge 
d\tchi_2\nn\\
&& 
 -\ft12{*d\varphi_1}\wedge d\varphi_1 
 -\ft12 e^{-2\varphi_1}\,{*dH_\3}\wedge dH_\3  \nn\\
&& -\ft12 e^{-\varphi_1}\, \Big[ 
 e^{\tvp_2-\varphi_3}\,{*\tbF}_1\wedge \tbF_1 + 
  e^{-\tvp_2+\varphi_3}\,{*\tbF}_2\wedge \tbF_2\nn\\
&& \qquad\qquad+
  e^{-\tvp_2-\varphi_3}\, {*\bF}^3\wedge\bF^3 +
  e^{\tvp_2+\varphi_3}\, {*\bF}^4\wedge\bF^4 \Big]\,,\label{lagtH}
\eea
where $\tF_1=d\tA_1$, $\tF_2=d\tA_2$, $F^3=dA^3$ and $F^4=dA^4$
are the ``raw'' field strengths,
\bea
\tbF_1&=&\tF_1-\tchi_2\, F^3\,,\qquad  \tbF_2=\tF_2 - \chi_3\, F^3\,,\nn\\
\bF^3&=& F^3\,,\qquad \bF^4= F^4 + \chi_3\, \tF_1 +\chi_2\, \tF_2 -
  \tchi_2\, \chi_3\, F^3\,,\label{tildeF}
\eea
are the ``dressed'' field strengths appearing in the kinetic terms in
(\ref{lagtH}), and
\be
H_\3 = dB_\2' -\tA_1\wedge d\tA_2 - A^3\wedge dA^4\,.
\ee

  Finally, we may dualise the 2-form potential $B_\2'$ to an axion $\chi_1$
by adding a Lagrangian multiplier term $\chi_1\, (dH_\3 + \tF_1\wedge
\tF_2 + F^3\wedge F^4)$ to (\ref{lagtH}), treating $H_\3$ as an 
independent field, and substituting its equation of motion $H_\3=
 -e^{2\varphi_1}\, {*d\chi_1}$ back into the Lagrangian.  This gives
the bosonic STU supergravity Lagrangian in the form
\bea
{\cal L} &=& R\,{*\oneone} -\ft12 \sum_{i=1,3}({*d\varphi_i}\wedge d\varphi_i
+ e^{2\varphi_i}\, {*d\chi_i}\wedge d\chi_i)
  -\ft12 {*d\tvp_2}\wedge d\tvp_2 -
  \ft12 e^{2\tvp_2}\, {*d\tchi_2}\wedge d\tchi_2 \nn\\
&& -\ft12 e^{-\varphi_1}\, \Big[ 
 e^{\tvp_2-\varphi_3}\,{*\tbF}_1\wedge \tbF_1 + 
  e^{-\tvp_2+\varphi_3}\,{*\tbF}_2\wedge \tbF_2 \label{lagt}\\
&&
\qquad\qquad
 + e^{-\tvp_2-\varphi_3}\, {*\bF}^3\wedge\bF^3 +
  e^{\tvp_2+\varphi_3}\, {*\bF}^4\wedge\bF^4 \Big]
+ \chi_1\, (\tF_1\wedge\tF_2 + F^3\wedge F^4)\,.\nn
\eea

\end{document}